\documentclass[]{aa}
\usepackage{graphicx,hyperref,txfonts}
\usepackage{amsbsy}
\usepackage{subcaption}

\begin{document}

\title{Generalised model-independent characterisation of strong gravitational lenses I: Theoretical foundations}
\titlerunning{Generalised model-independent characterisation of strong gravitational lenses I}
\author{J. Wagner}
\institute{Universit\"at Heidelberg, Zentrum f\"ur Astronomie, Institut f\"ur Theoretische Astrophysik, Philosophenweg 12, 69120 Heidelberg, \\ Heidelberg Institute for Theoretical Studies, 69118 Heidelberg, Germany\\
 \email{j.wagner@uni-heidelberg.de}}
\date{Received XX; accepted XX}

\abstract{We extend the model-independent approach to characterise strong gravitational lenses of \cite{bib:Wagner} to its most general form to leading order by using the orientation angles of a set of multiple images with respect to their connection line(s) in addition to the relative distances between the images, their ellipticities and time-delays. For two symmetric images that straddle the critical curve, the orientation angle additionally allows to determine the slope of the critical curve and  a second (reduced) flexion coefficient at the critical point on the connection line between the images. It also allows to drop the symmetry assumption that the axis of largest image extension is orthogonal to the critical curve. For three images almost forming a giant arc, the degree of assumed image symmetry is also reduced to the most general case, allowing to describe image configurations for which the source need not be placed on the symmetry axis of the two folds that unite at the cusp. For a given set of multiple images, we set limits on the applicability of our approach, show which information can be obtained in cases of merging images, and analyse the accuracy achievable due to the Taylor expansion of the lensing potential for the fold case on a galaxy cluster scale NFW-profile, a fold and cusp case on a galaxy cluster scale SIE-profile, and compare the generalised approach with the one of \cite{bib:Wagner}. The position of the critical points is reconstructed with less than 5'' deviation for multiple images closer to the critical points than 30\% of the (effective) Einstein radius and the slope of the critical curve at a fold and its shape in the vicinity of a cusp deviate less than 20\% from the true values for distances of the images to the critical points less than 15\% of the (effective) Einstein radius.}


\maketitle

%%%%%%%%%%%%%%%%%
\section{Introduction and motivation}
\label{sec:introduction}

While information from the second order multipole moments of weakly distorted sources is routinely extracted in weak lensing, \cite{bib:Bartelmann, bib:KSB, bib:Refregier, bib:Seitz}, we show that these multipole moments also yield vital information about the lensing potential when being extracted from multiple images in the strong lensing regime. We generalise the approach developed in \cite{bib:Wagner} by including the orientation angle of the semi-major axis of the images into our set of observables and thereby reconstruct the critical curve in the vicinity of multiple images for image configurations of two images that straddle the critical curve and for three images almost forming a giant arc without assuming any lens model. 
We aim at demonstrating that a lot of information contained in multiply-imaged source observations is still unused and can be easily employed to retrieve local, general properties of the lensing potential without the need of highly degenerate model fitting or ad-hoc fine-tuning by adding substructure.

The paper is organised as follows: 

Sect.~\ref{sec:equations} briefly reviews the method as developed in \cite{bib:Wagner} and presents the definitions of observables and assumptions required before listing the characteristics of the lensing potential that can be inferred from the observables. 

In Sect.~\ref{sec:accuracy_and_limits}, we analyse the influence of the Taylor expansion on the reconstruction of the critical curve and the ratios of potential derivatives and calculate which information can be retrieved in the case of overlapping images. 

As a simple test case, we show a simulated galaxy cluster consisting of an NFW-profile as gravitational lens and a Sérsic profile for the sources in Sect.~\ref{sec:simulation}. We discuss the results from the simulation and show the accuracy limits of the Taylor expansion under the influence of the finite pixel resolution and  the approximation of the image shape as quadrupole moment. Subsequently, we show the reconstruction of the critical curve and the potential derivatives obtained by the generalised approach for multiple images at a fold and cusp in a simulated galaxy cluster scale singular isothermal ellipse (SIE) and compare the results to the ones obtained without taking into account the image orientations.

Sect.~\ref{sec:conclusion} summarises the results and gives an outlook to the next steps necessary to make our approach applicable to observations.

Being the first part of a series of papers, this work introduces the general formalism and the complete set of observables for the leading order Taylor expansion. In subsequent papers, we will analyse realistic simulations like the HERA cluster, \cite{bib:Meneghetti}, and investigate the influence of the source properties, as well as biases and uncertainties introduced by the observation process in detail.

%%%%%%%%%%%%%%%%%%

\begin{figure*}[h!]
\centering
\begin{subfigure}{0.49\textwidth}
  \centering
 \includegraphics[width=0.7\linewidth]{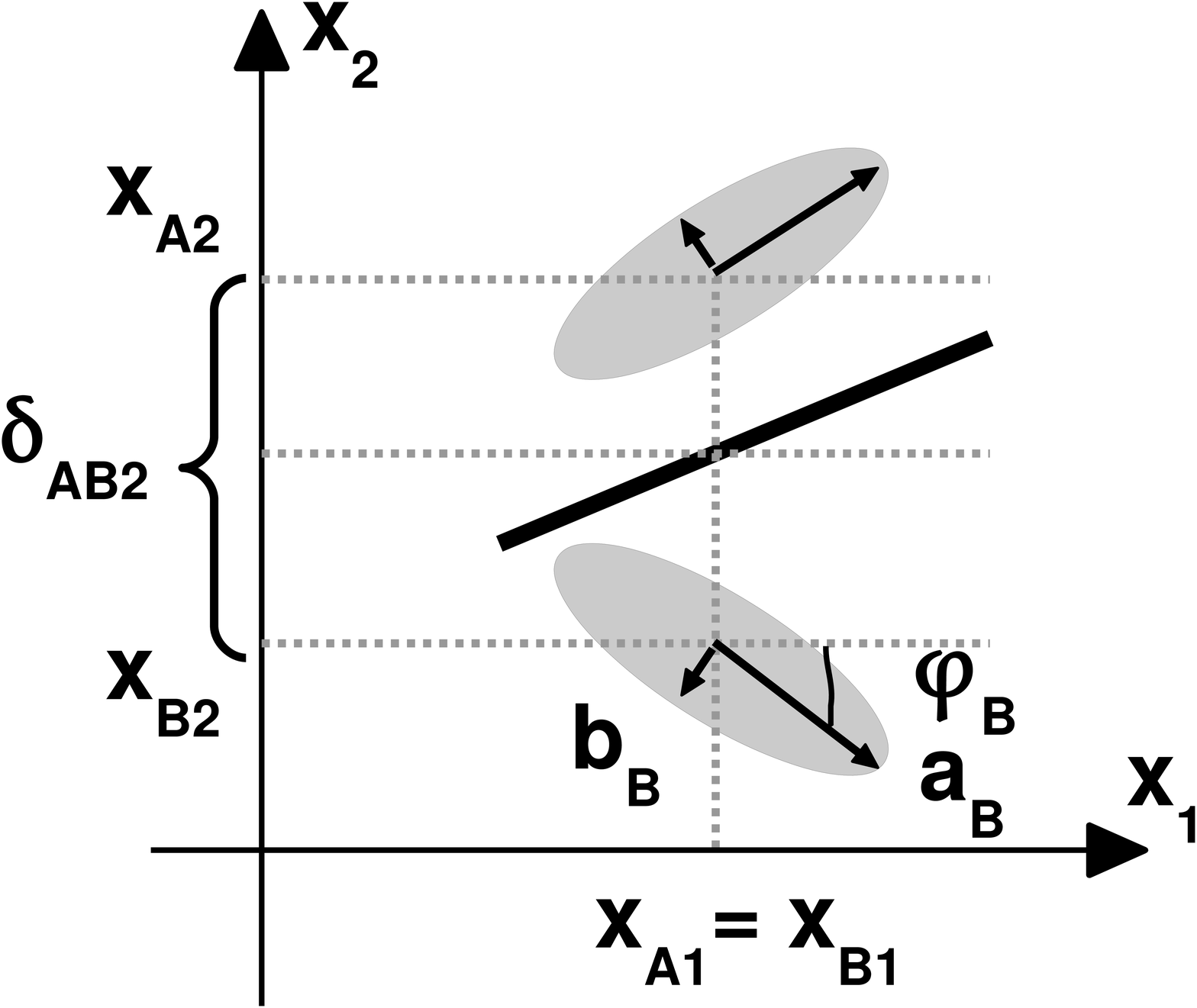}
 \caption{}
 \end{subfigure}
 \begin{subfigure}{0.49\textwidth}
   \centering
  \includegraphics[width=0.63\linewidth]{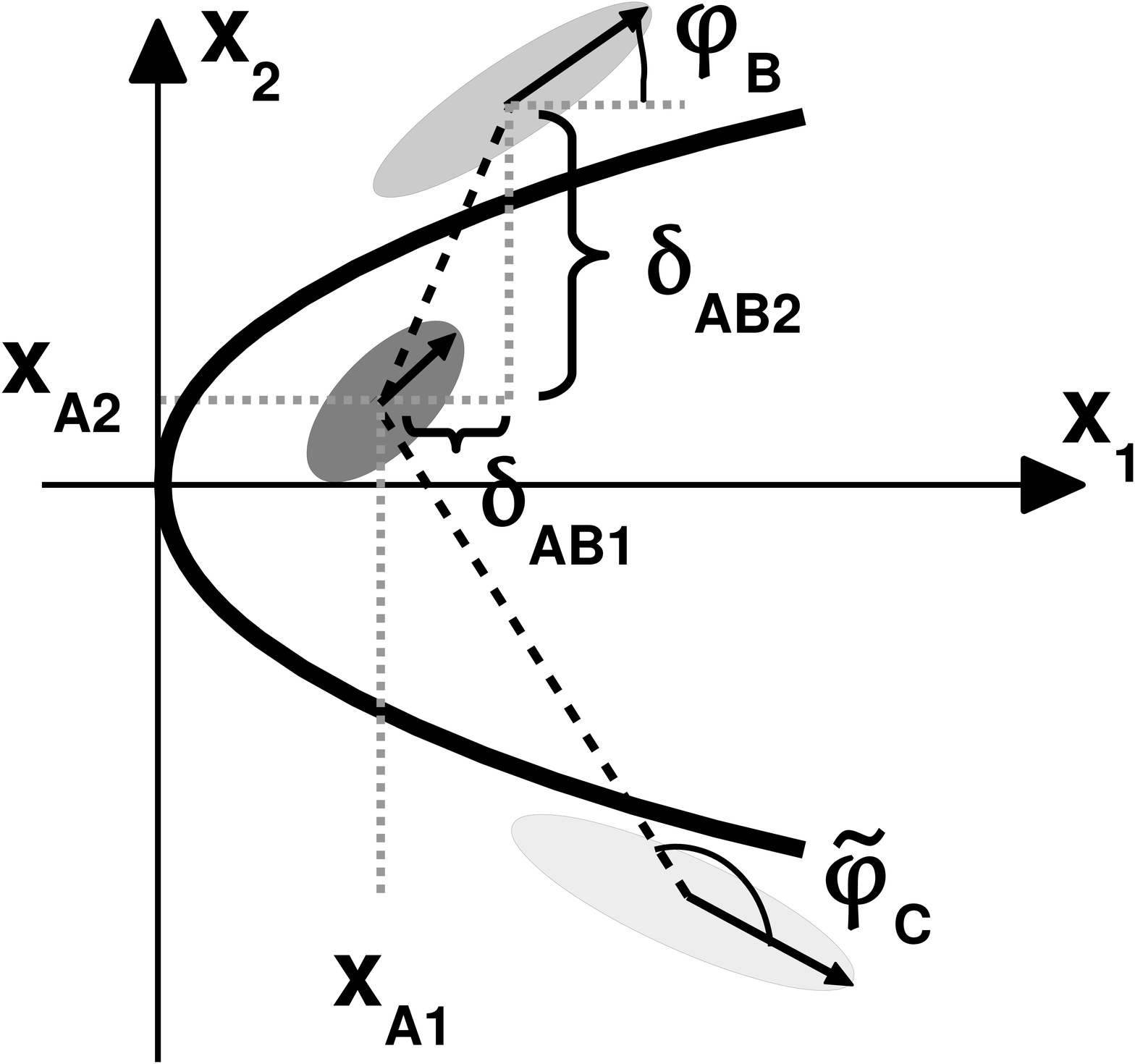}
  \caption{}
 \end{subfigure}
    \caption{Coordinate systems and definitions of observables for two images close to a fold critical point (a), three images close to a cusp critical point (b). The thick black line in (a) and parabola in (b) show the approximation of the critical curve in the vicinity of the images. The fold critical point under analysis in (a) is located at the crossing of the critical curve and the connection line of image $A$ and $B$, the cusp critical point in (b) is the vertex of the parabola.}
\label{fig:coordinate_systems}
\end{figure*}

\section{Observables and (generalised) approach}
\label{sec:equations}

\subsection{Definitions and assumptions}

We expand the gravitational lensing potential $\phi(\boldsymbol{x},\boldsymbol{y})$ in a Taylor series around a point on the critical curve, $\boldsymbol{x}_{0} \in \mathbb{R}^2$, and its respective point on the caustic, $\boldsymbol{y}_{0} \in \mathbb{R}^2$, as described in detail in \cite{bib:Wagner} and transform into a coordinate system in which 
\begin{equation}
\left\{ \boldsymbol{x}_{0}, \boldsymbol{y}_{0}\right\} = \left\{ (0,0),(0,0)\right\}\;, \quad  \phi_1^{(0)} = \phi_2^{(0)} = \phi_{12}^{(0)} = \phi_{22}^{(0)} = 0
\label{eq:coordinate_system}
\end{equation}
hold. A subscript $i$ of $\phi$ denotes the partial derivative in $x_i$-direction, $i=1,2$, and the superscript $(0)$ indicates that this derivative is evaluated at the singular point. We then employ as observables
\begin{itemize}
\item the time delay $t_{ij} = t_{i} - t_{j}$ between the arrival times of two images $i$ and $j$,
\item the relative coordinate distances $\delta_{ij1} = x_{i1} - x_{j1}$ and $\delta_{ij2} = x_{i2} - x_{j2}$ between two images $i$, $j$ along the axes of the coordinate system defined by Eq.~\eqref{eq:coordinate_system},
\item the axis ratio $r_i = b_i/a_i$ between the semi-minor axis $b_i$ and the semi-major axis $a_i$ of the quadrupole moment of image $i$,
\item the orientation angle $\varphi_i$ between the semi-major axis of image $i$ and the $x_1$-axis of the coordinate system chosen by Eq.~\eqref{eq:coordinate_system} (we will also use $\lambda_i \equiv \cos(\varphi_i)/\sin(\varphi_i)$), and
\item the flux ratios $\mu_i/\mu_j$ between image $i$ and $j$, representing the magnification ratios.
\item If time delay information is available, we also require the redshift of the source, $z_\mathrm{s}$, and the lens, $z_\mathrm{d}$, to calculate the angular diameter distances between the observer and the source, $D_\mathrm{s}$, between the observer and the lens, $D_\mathrm{d}$, and between the lens and the source, $D_\mathrm{ds}$, which yield the distance factor $\Gamma_\mathrm{d} \equiv (D_\mathrm{d} D_\mathrm{s})/(D_\mathrm{ds}) (1+z_\mathrm{d})$.
\end{itemize}

\noindent
Fig.~\ref{fig:coordinate_systems} illustrates the definitions of observables and choices of coordinate systems for the fold configuration consisting of two images (a) and for the cusp configuration consisting of three images (b). 

We assume that the images are located so close to the critical curve that intrinsic ellipticities are negligible compared to the distortions caused by lensing. This assumption is justified, as it can be shown that the observables of the images are completely dominated by the lens mapping on the critical curve and that the influence of the lens decreases for increasing distance from the critical curve (see Appendix~\ref{app:source_properties} for details). The analysis of the intrinsic properties and higher order effects will be treated separately.

\subsection{Resulting (ratios of) potential derivatives at folds}

We assume that a source is close to a fold singular point of the caustic, such that it is mapped to two images in the lens plane, denoted by $A$ and $B$. They straddle the critical curve, as shown in Fig.~\ref{fig:coordinate_systems} (right). We assume, without loss of generality, that $A$ has positive and $B$ negative parity. As derived in more detail in Appendix~\ref{app:fold_derivation}, we obtain from the observables of these two images
\begin{align}
\phi_{222}^{(0)} &= \dfrac{12 c t_{AB}}{\Gamma_\mathrm{d} \delta_{AB2}^3}\;, \label{eq:fold1}\\
\dfrac{\phi_{222}^{(0)}}{\phi_{11}^{(0)}} &= \dfrac{2\left( r_i + \lambda_i^2\right)}{\delta_{AB2}\left( r_i \lambda_i^2 +1\right)}\;, \quad &i = A,B \label{eq:fold2} \\
\dfrac{\phi_{122}^{(0)}}{\phi_{11}^{(0)}} &= \dfrac{2\lambda_A(r_A -1)}{\delta_{AB2}\left( r_A \lambda_A^2 +1\right)} = - \dfrac{2\lambda_B(r_B -1)}{\delta_{AB2}\left( r_B \lambda_B^2 +1\right)}\;,\label{eq:fold3} \\
\dfrac{\phi_{122}^{(0)}}{\phi_{222}^{(0)}} &= - \dfrac{\lambda_A (r_A -1)}{r_A + \lambda_A^2 } =  \dfrac{\lambda_B (r_B -1)}{r_B + \lambda_B^2 }\;. \label{eq:fold4} 
\end{align}
For $|\varphi_i| = \pi/2$, Eq.~\eqref{eq:fold2} becomes  Eq.~10 of \cite{bib:Wagner}. If the time delay has been measured, Eqs.~\eqref{eq:fold2} and \eqref{eq:fold3} can be solved for $\phi_{11}^{(0)}$ and $\phi_{122}^{(0)}$. Although Eq.~\eqref{eq:fold4} can be determined from Eqs.~\eqref{eq:fold2} and \eqref{eq:fold3}, we explicitly show this ratio as it is equal to the slope of the critical curve at $\boldsymbol{x}_{0}$. For $|\varphi_i| = \pi/2$, it becomes zero so that images and critical curve are orthogonal to each other.
Significant deviations between the ratios obtained from $A$ and $B$ can arise, if the images are too far from the critical curve for the assumption $|\mu_A| = |\mu_B|$ to hold or a measurement bias (e.g.\ due to microlensing or dust extinction) has not been taken into account yet.

The positions of $A$ and $B$ with respect to the critical point under consideration are given by
\begin{align}
x_{A1} = x_{B1} = 0 \;, \quad x_{A2} = - x_{B2} = \dfrac{\delta_{AB2}}{2} \;.
\end{align}

\subsection{Resulting (ratios of) potential derivatives at cusps}

\begin{figure*}[ht!]
\centering
\begin{subfigure}{0.29\textwidth}
  \centering
  \includegraphics[width=0.7\linewidth]{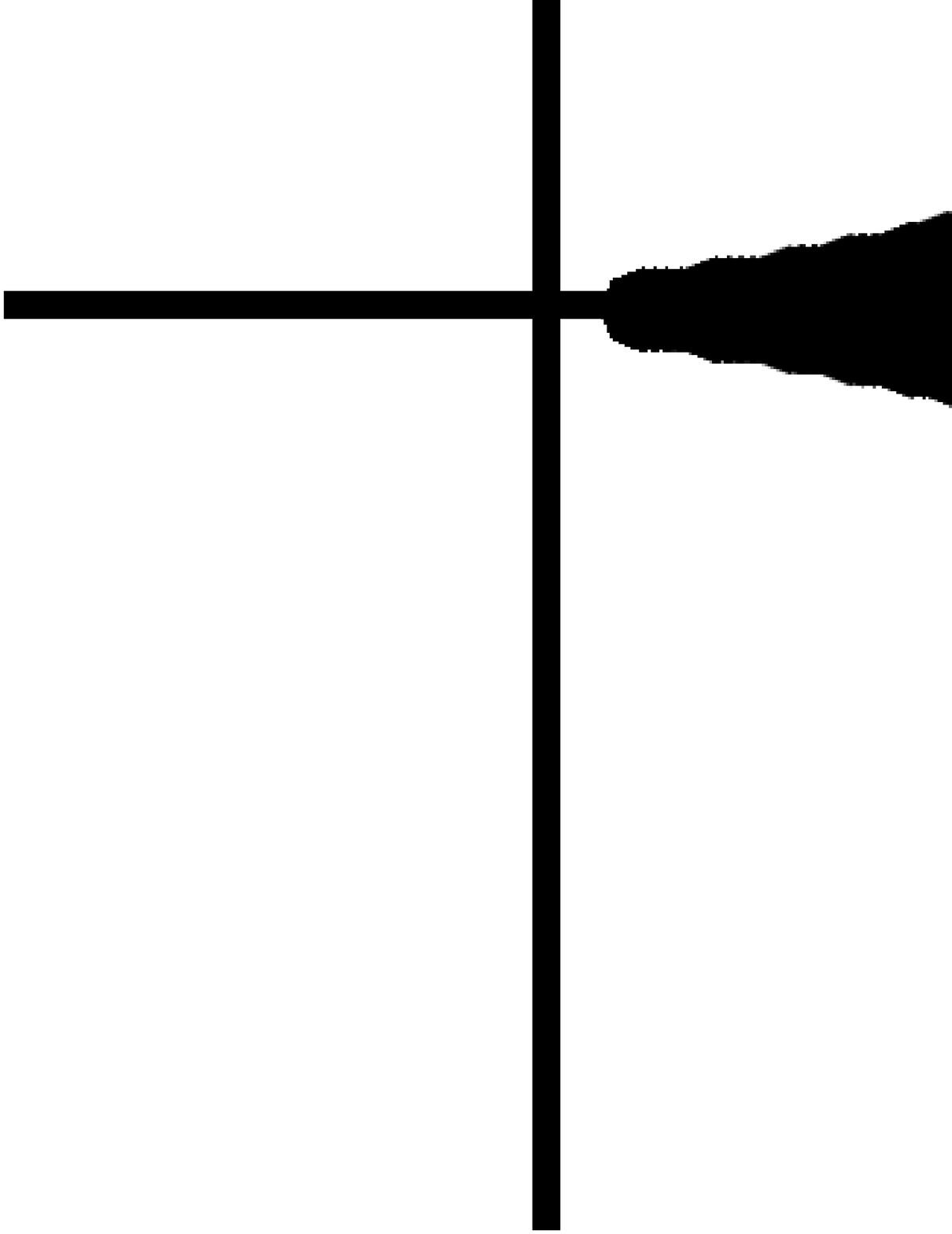}
  \caption{}
  \end{subfigure}
  \begin{subfigure}{0.29\textwidth}
  \centering
  \includegraphics[width=0.7\linewidth]{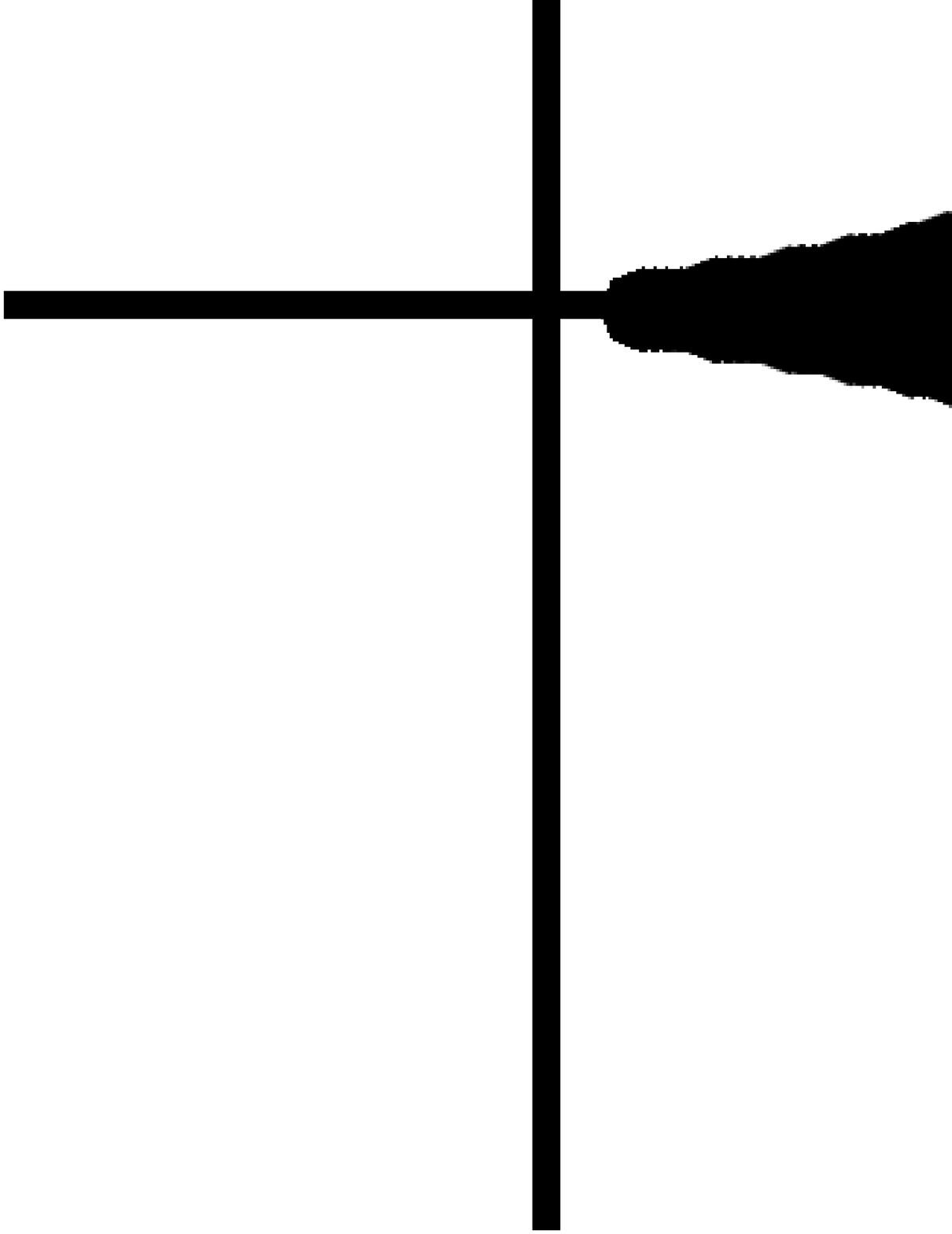}
  \caption{}
  \end{subfigure}
  \begin{subfigure}{0.29\textwidth}
  \centering
  \includegraphics[width=0.7\linewidth]{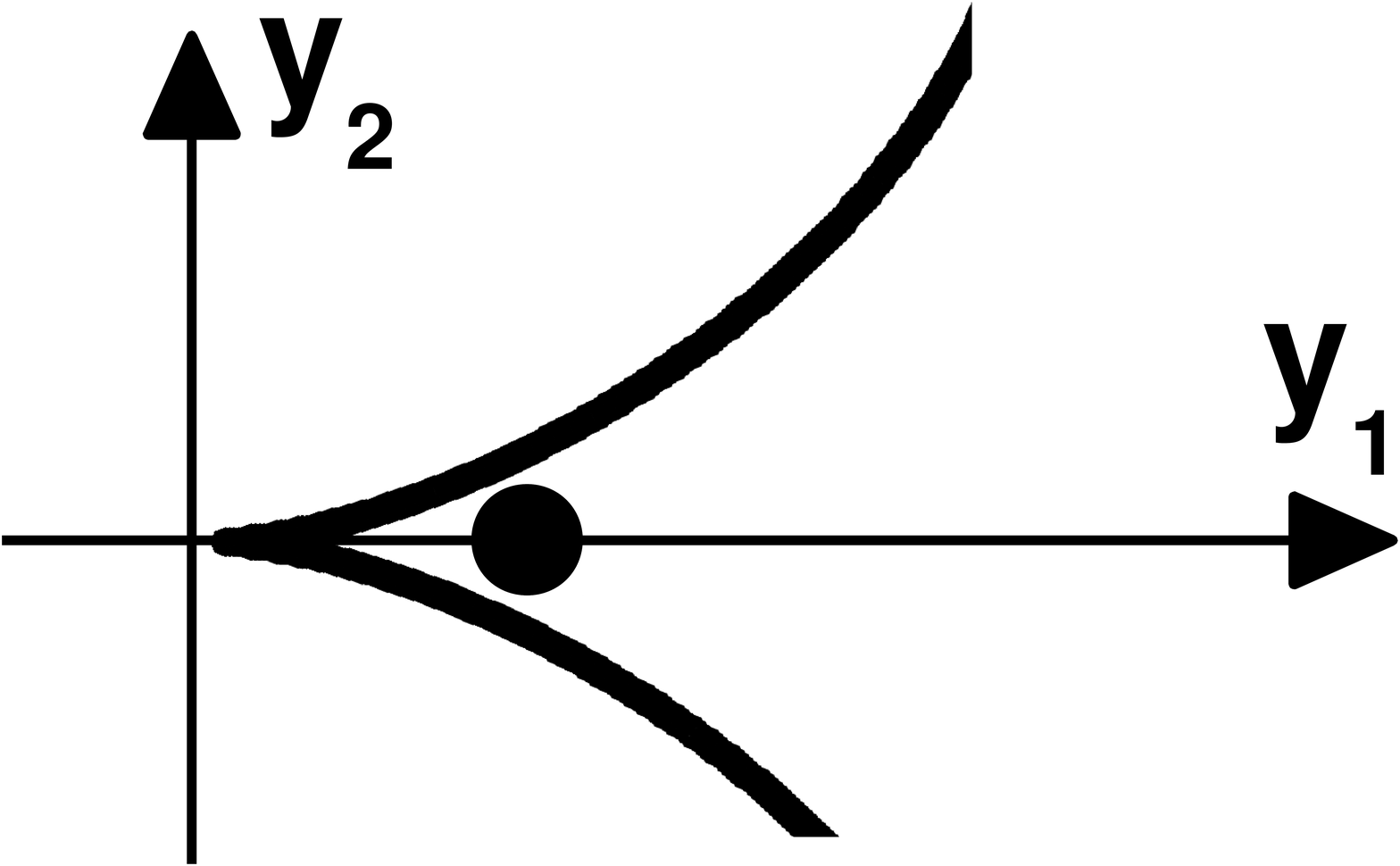}
  \caption{}
  \end{subfigure}
  \begin{subfigure}{0.29\textwidth}
  \centering
  \includegraphics[width=0.7\linewidth]{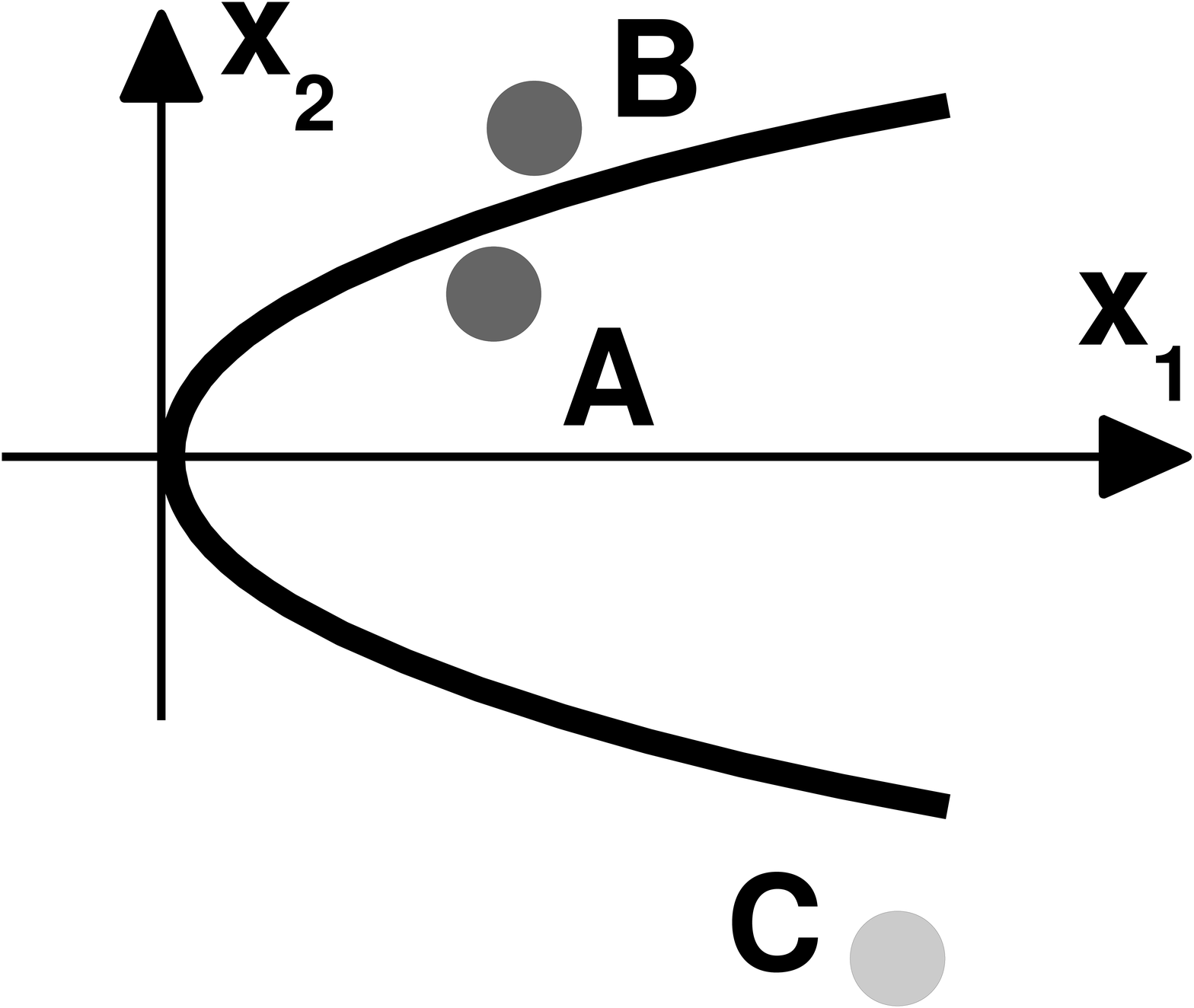}
  \caption{}
  \end{subfigure}
  \begin{subfigure}{0.29\textwidth}
  \centering
  \includegraphics[width=0.7\linewidth]{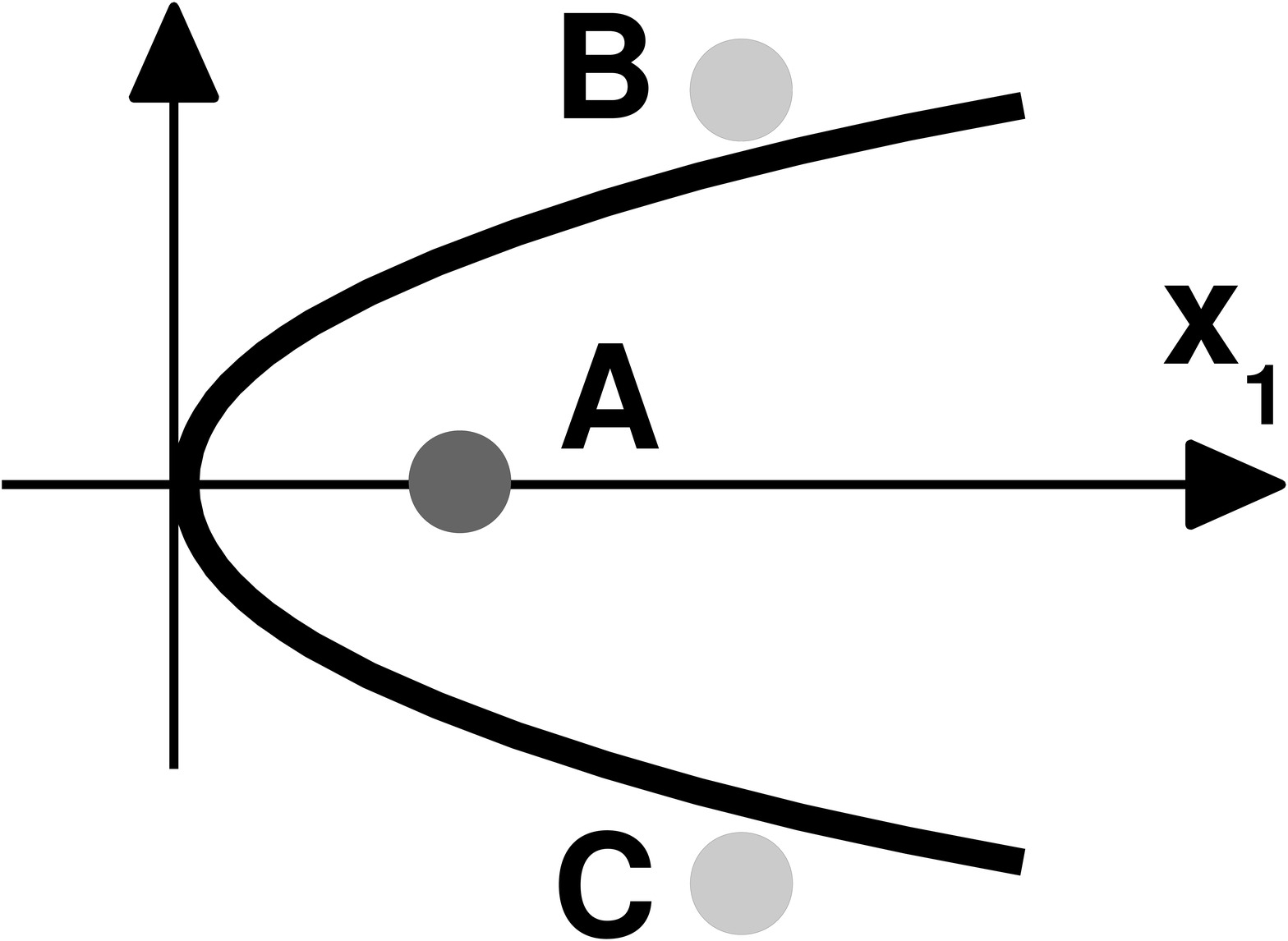}
  \caption{}
  \end{subfigure}
  \begin{subfigure}{0.29\textwidth}
  \centering
  \includegraphics[width=0.7\linewidth]{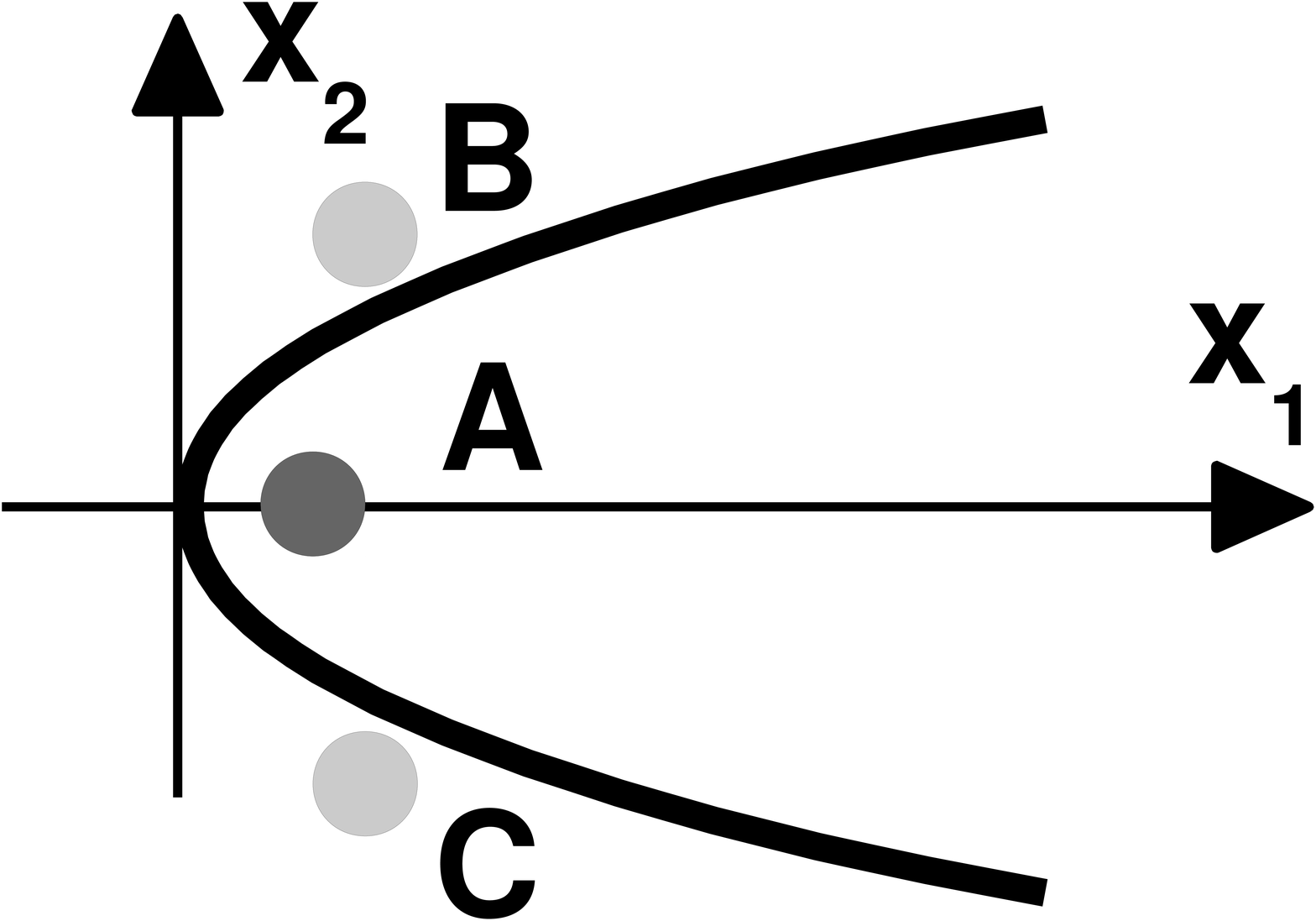}
  \caption{}
  \end{subfigure}
    \caption{Observing three images (d)-(f), an analysis of their observables determines whether to apply the fold or cusp equations and where the source is located with respect to the folds and cusp (a)-(c): a fold configuration (a) and (e) shows two highly symmetric images, $A$, $B$, with larger separation to the third image, $C$, less magnified than $A$, $B$. In a cusp configuration (b) and (e), (c) and (f), $A$ is highly magnified compared to $B$ and $C$ because it can be considered as an overlay of two images generated by lensing at the upper and the lower fold which unite at the cusp point. Image shapes are neglected in these graphs but high and low absolute values of magnifications are indicated by dark and light grey values, respectively.}
\label{fig:fold_vs_cusp}
\end{figure*}

The images $A$, $B$, and $C$ close to a cusp cannot be treated on equal footing with each other because they have different distances to the critical point. To obtain the best Taylor-approximation of the (ratios of) potential derivatives, observables from $A$ (which is closest to the cusp, i.e.\ the expansion point of the Taylor series) are used in combination with those from $B$ or $C$. Combinations of observables from $B$ with $C$ are not employed, as they yield less accurate results because $B$ and $C$ are farther from the cusp, as will be shown in Sects.~\ref{sec:sie_cusp} and \ref{sec:sie_asymmetric_cusp}. They are also linearly dependent on the other combinations and therefore do not contain further information.
As detailed in Appendix~\ref{app:cusp_derivation}, we obtain for the (ratios of) potential derivatives
\begin{align}
\phi_{11}^{(0)} &= \dfrac{-2c t_{AB}}{\Gamma_\mathrm{d}} \left(\boldsymbol{\delta}^\top \tilde{M}_A \boldsymbol{\delta} - \dfrac{\phi_{122}^{(0)}}{\phi_{11}^{(0)}} \delta_{AB1} \delta_{AB2}^2 + \dfrac{1}{12} \dfrac{\phi_{2222}^{(0)}}{\phi_{11}^{(0)}} \delta_{AB2}^4 \right)^{-1} \;, \label{eq:cusp1} \\
\dfrac{\phi_{122}^{(0)}}{\phi_{11}^{(0)}} &= \dfrac{1}{\delta_{AB2}}\left(  \dfrac{\lambda_A (r_A - 1) }{r_A \lambda_A^2 + 1} - \dfrac{\lambda_B (r_B - 1) }{r_B \lambda_B^2 + 1} \right) \;, \label{eq:cusp2} \\
\dfrac{\phi_{2222}^{(0)}}{\phi_{11}^{(0)}} &=  \dfrac{-6}{\delta_{AB2}^{2}}\left( \dfrac{\delta_{AB1}}{\delta_{AB2}} \dfrac{\lambda_B (r_B - 1) }{r_B \lambda_B^2 + 1} \stackrel{-}{+} \dfrac{r_A + \lambda_A^2 }{r_A \lambda_A^2 + 1} \right) \;, \label{eq:cusp3}
\end{align}
and for the position of $A$ with respect to $\boldsymbol{x}_{0} = (0,0)$
\begin{align}
x_{A1} &= - \left( \dfrac{\phi_{122}^{(0)}}{\phi_{11}^{(0)}}\right)^{-1} \left(\dfrac12 \dfrac{\phi_{2222}^{(0)}}{\phi_{11}^{(0)}} x_{A2}^2 \stackrel{-}{+}  \dfrac{r_A + \lambda_A^2 }{r_A \lambda_A^2 + 1} \right) \;, \label{eq:cusp_coords1} \\
x_{A2} &= \dfrac{\delta_{AB2}}{2} + \dfrac{\left(1 - \tfrac{\mu_A}{\mu_B}\right) \det(\tilde{M}_A) - \tfrac{\phi_{122}^{(0)}}{\phi_{11}^{(0)}}\delta_{AB1}}{2\delta_{AB2} \left(\tfrac12\tfrac{\phi_{2222}^{(0)}}{\phi_{11}^{(0)}} - \left( \tfrac{\phi_{122}^{(0)}}{\phi_{11}^{(0)}}\right)^2  \right)} \;. \label{eq:cusp_coords2}
\end{align}
with
\begin{equation}
\boldsymbol{\delta} = \left( \begin{matrix} \delta_{AB1} \\ \delta_{AB2}\end{matrix} \right) \;, \quad 
\tilde{M}_A = \left(  \begin{matrix} 1 & \dfrac{\lambda_A (r_A - 1) }{r_A \lambda_A^2 + 1} \\ \dfrac{\lambda_A (r_A - 1) }{r_A \lambda_A^2 + 1} & \stackrel{-}{+} \dfrac{r_A + \lambda_A^2 }{r_A \lambda_A^2 + 1} \end {matrix} \right) \;.
\label{eq:det_MA}
\end{equation}
The upper signs indicate a positive cusp configuration with positive $\phi_{2222}^{(0)}/\phi_{11}^{(0)}$ and $A$ having negative parity, the lower signs are valid for a negative cusp configuration with negative $\phi_{2222}^{(0)}/\phi_{11}^{(0)}$ and a positive parity $A$ (s.\ \cite{bib:Petters}).

The slope $m_\mathrm{c}$ of the parabola that approximates the critical curve in the vicinity of the images and has its vertex in $\boldsymbol{x}_{0} = (0,0)$ is then given by
\begin{equation}
m_\mathrm{c} = - \left(\dfrac12 \dfrac{\phi_{2222}^{(0)}}{\phi_{11}^{(0)}} - \left( \dfrac{\phi_{122}^{(0)}}{\phi_{11}^{(0)}}\right)^2\right) \left(\dfrac{\phi_{122}^{(0)}}{\phi_{11}^{(0)}}\right)^{-1} \;. \label{eq:parabola_slope}
\end{equation}

Given time delay information, the cusp type can be determined. Fixing the sign of the $x_1$-coordinates of $A$, $B$ and $C$,  all signs of the potential derivatives are also fixed then. 

Replacing the observables from $B$ by those from $C$, we determine the ratios of potential derivatives  from the combination $A$ and $C$. For each combination of images, $A$ and $B$ and $A$ and $C$ we thus obtain (ratios of) potential derivatives and a parabolic approximation of the critical curve in the vicinity of all three images. If the results using the pair ($A$, $B$) and using the pair ($A$, $C$) in Eqs.~\eqref{eq:cusp1}--\eqref{eq:parabola_slope} coincide within the measurement uncertainty and the bias due to the Taylor expansion, the two folds that unite to the cusp point are symmetric. Any significant deviation between the results can be caused by a so-far unaccounted-for measurement bias in an observable or an asymmetry in the mass density causing this cusp in the lens map, so that the Taylor approximation turns inapplicable.

%%%%%%%%%%%%%%%%%%%%%%%
\section{Accuracy and limits}
\label{sec:accuracy_and_limits}

\subsection{Biases due to Taylor expansions}
In general, as already stated in \cite{bib:Wagner}, the accuracy of the approach depends on the terms of the Taylor expansion of the lensing potential that are taken into account. In this leading order approach, we neglect $\phi_{111}^{(0)}$ when establishing the lensing equations for the fold and assume that $\phi_{112}^{(0)} x_{1} \ll \phi_{11}^{(0)}$. For the cusp, we omit  $\phi_{111}^{(0)}, \phi_{112}^{(0)}$ and the fourth order derivatives except $\phi_{2222}^{(0)}$. Furthermore, we neglect the flexion coefficient at the location of the image, assuming that it is much smaller than the distortions caused by convergence and shear\footnote{In the detailed derivations in Appendices~\ref{app:fold_derivation} and \ref{app:cusp_derivation}, we show that no flexion coefficients at the image positions occur in the fold case and we indicate where we use this assumption in the cusp case.}. 

Secondly, the accuracy of the results is determined by the relative distances between the images and the expansion point of the Taylor series, i.e.\ the critical point. As the approach also includes shape information, the quadrupole moment may be biased by higher order shape distortions.

In order to investigate the impact of these approximations and assumptions on the accuracy, simulations are set up, as detailed in Sect.~\ref{sec:simulation}. Furthermore, we will investigate the impact of including the orientation angle compared to the approach of \cite{bib:Wagner} without image orientations.

\subsection{Analysis of observed triple-image configurations}
\label{sec:cusp_fold}

To leading order Taylor expansion, the two images of a fold configuration are assumed to be so close to the critical curve that the images have a high symmetry with similiar shapes and equal absolute fluxes. Hence, finding an image configuration with observables that fulfil this criterion, we apply the equations for the fold case. In cases where a third image is close by, we can additionally apply the equations for the cusp configuration to retrieve information about the cusp critical point. Then, two approximations to the critical curve -- a linear one from the fold evaluation and a parabolic one from the cusp evaluation -- in the vicinity of the two images with high symmetry are obtained. In the following, we will derive the relationship between them.

The (ratios of) potential derivatives obtained at the cusp point will be subject to a higher deviation due to the Taylor expansion because a ``genuine'' cusp configuration occurs when the source is close to both folds (and thus close to the cusp) that the magnifications of the two outer images become equal and about half of the magnification of the inner one. Fig.~\ref{fig:fold_vs_cusp} sketches the situations qualitatively. 

\begin{figure*}[ht!]
\centering
\begin{subfigure}{0.29\textwidth}
  \centering
  \includegraphics[width=0.7\linewidth]{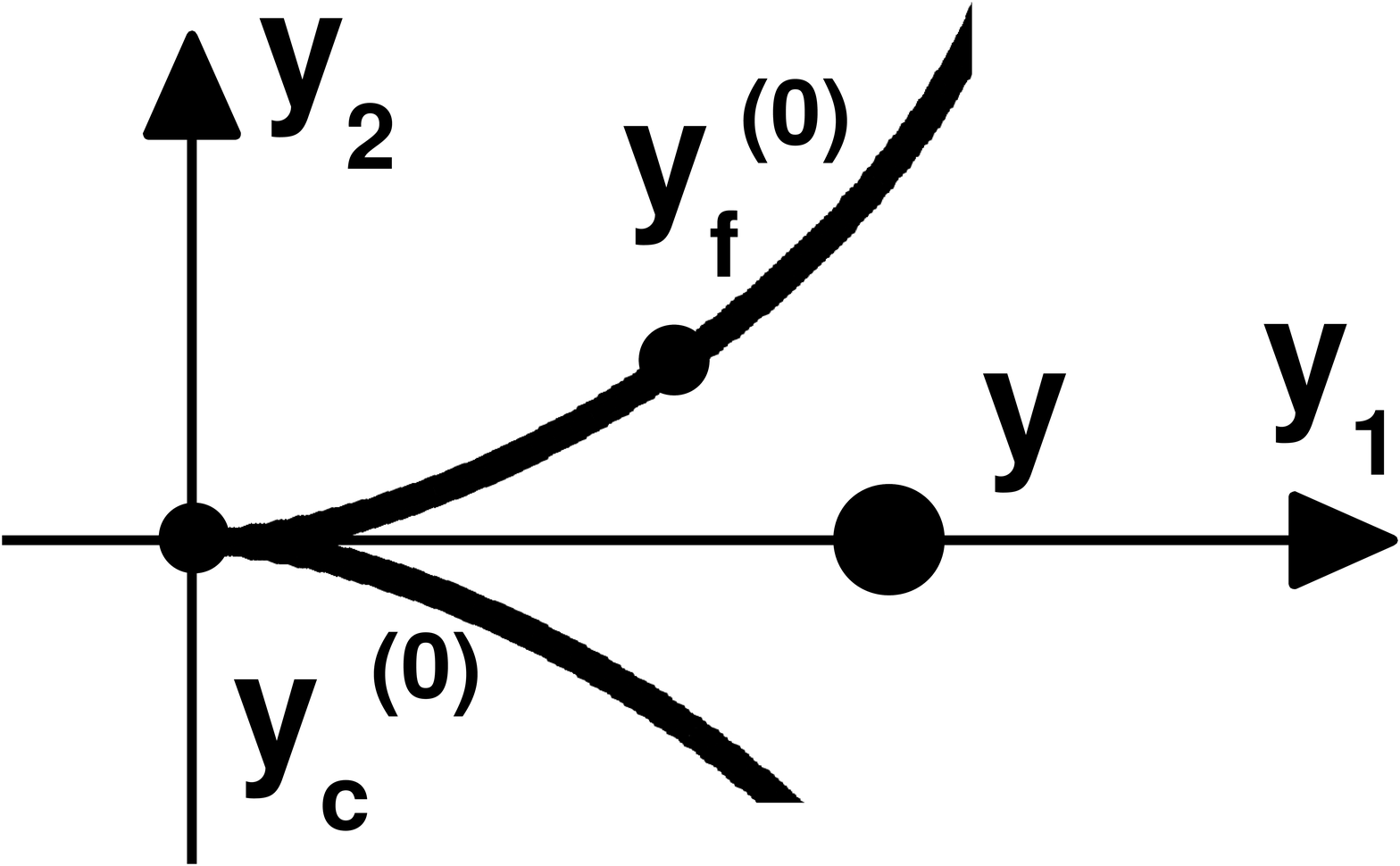}
  \caption{}
  \end{subfigure}
  \begin{subfigure}{0.29\textwidth}
  \centering
  \includegraphics[width=0.7\linewidth]{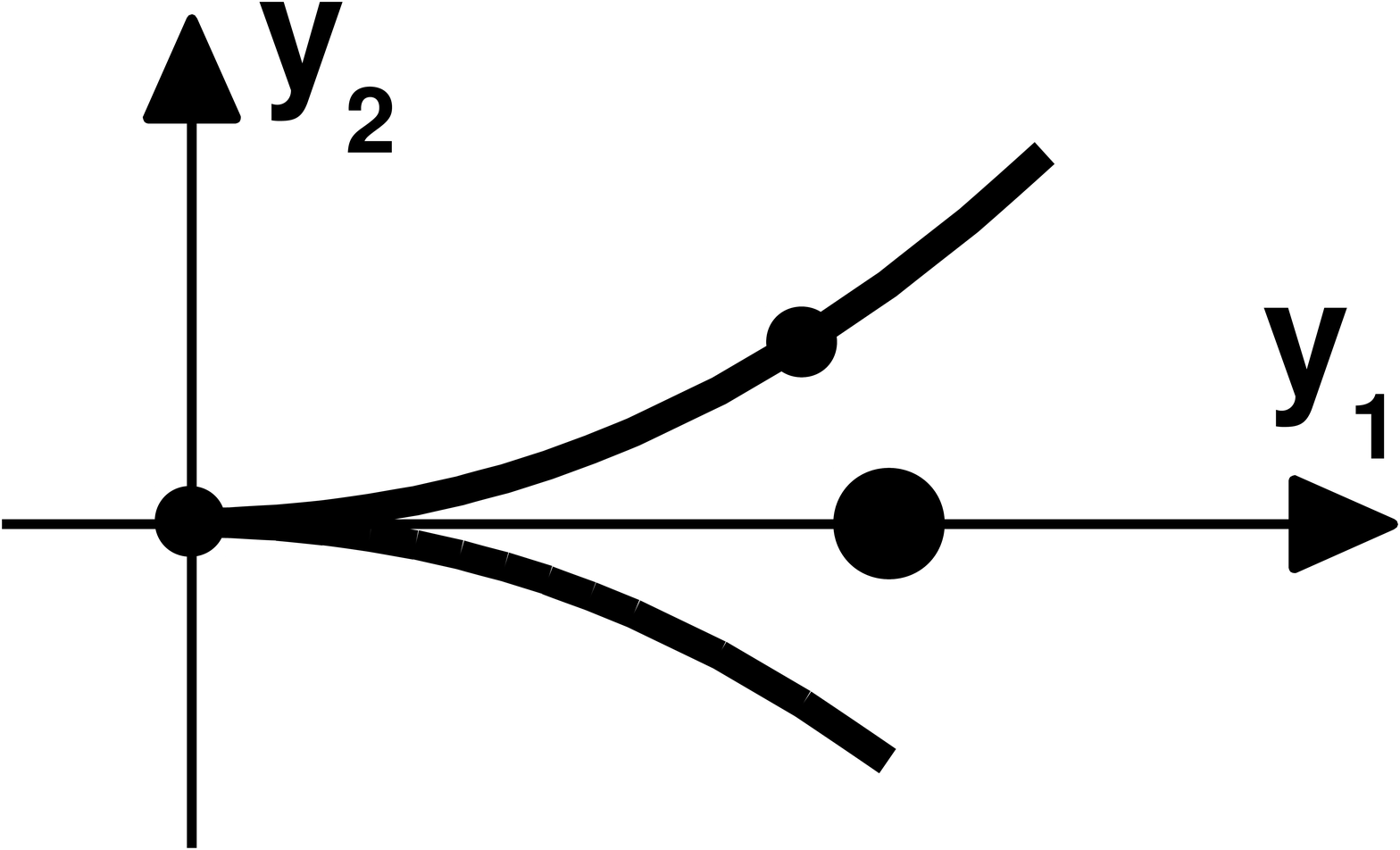}
  \caption{}
  \end{subfigure}
  \begin{subfigure}{0.29\textwidth}
  \centering
  \includegraphics[width=0.7\linewidth]{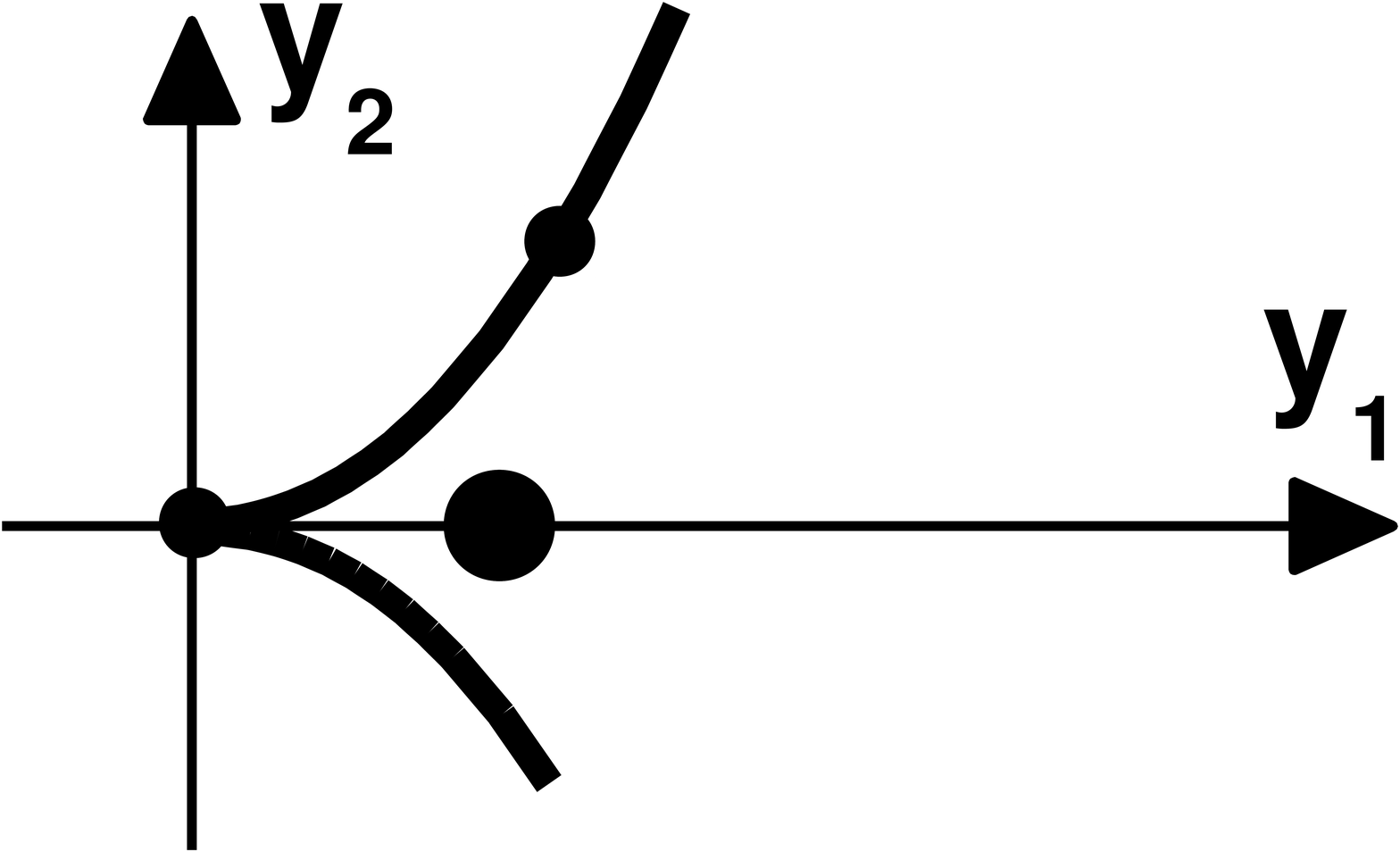}
  \caption{}
  \end{subfigure}
  \begin{subfigure}{0.29\textwidth}
  \centering
  \includegraphics[width=0.7\linewidth]{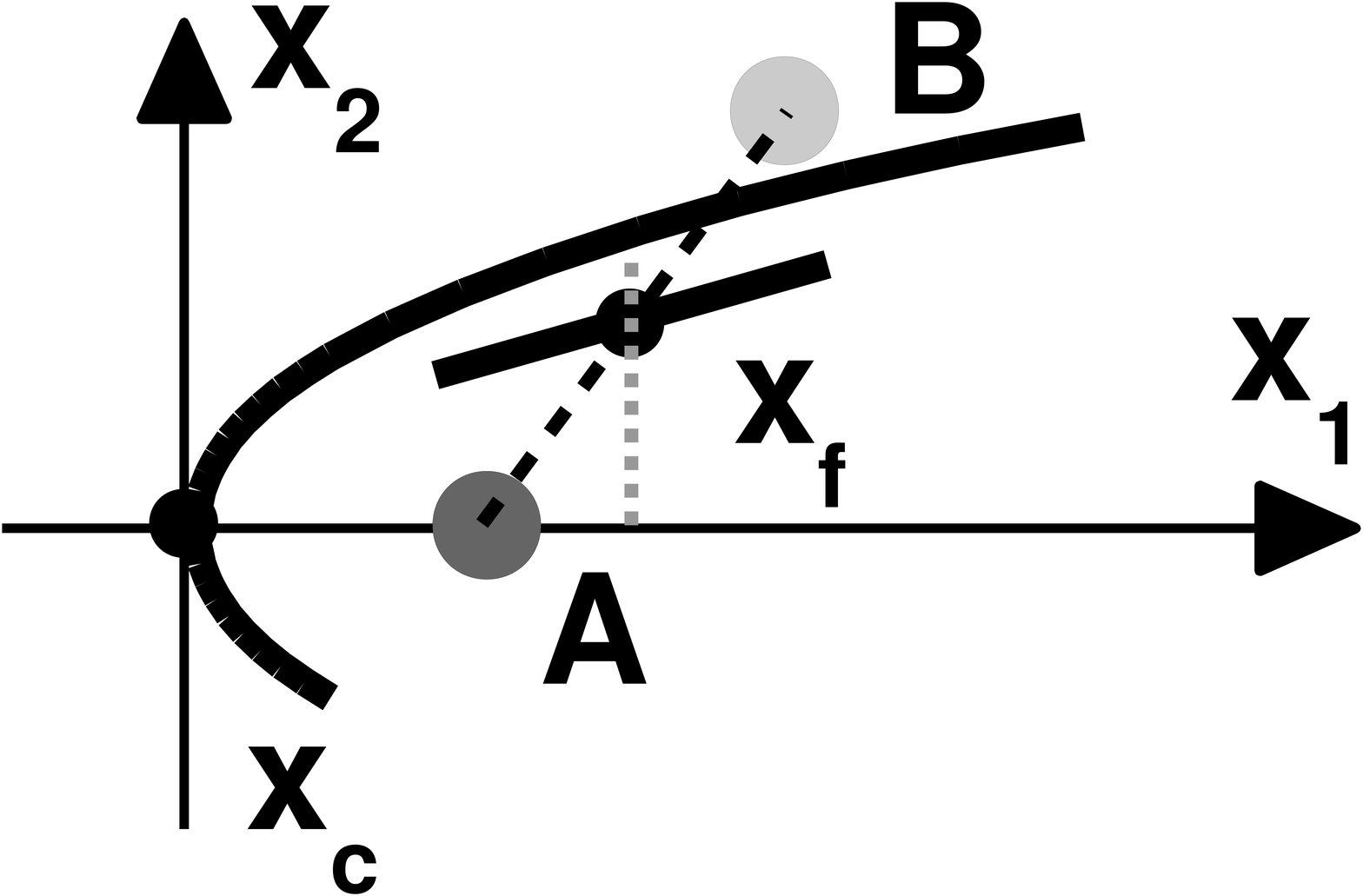}
  \caption{}
  \end{subfigure}
  \begin{subfigure}{0.29\textwidth}
  \centering
  \includegraphics[width=0.7\linewidth]{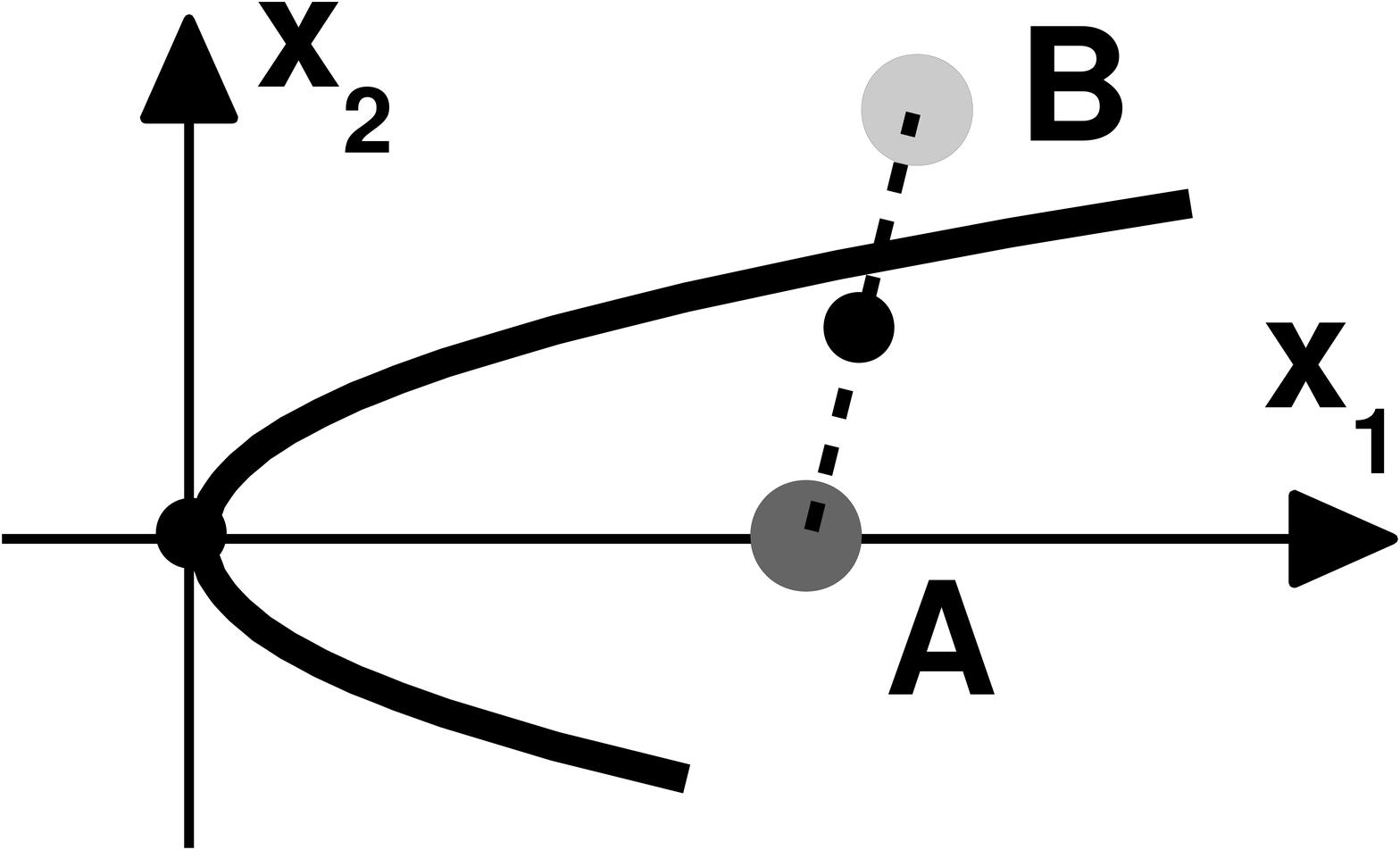}
  \caption{}
  \end{subfigure}
  \begin{subfigure}{0.29\textwidth}
  \centering
  \includegraphics[width=0.7\linewidth]{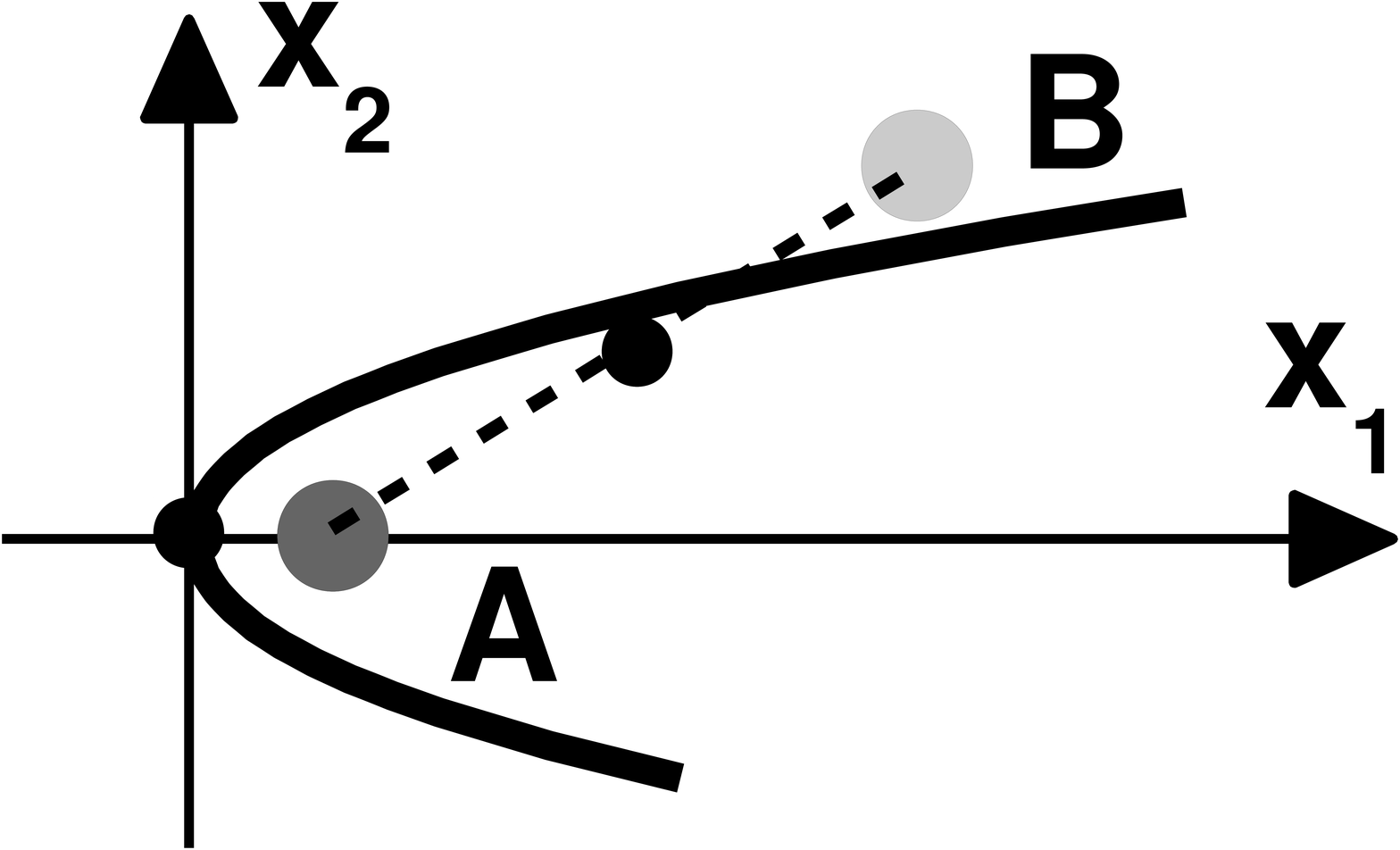}
  \caption{}
  \end{subfigure}
    \caption{Comparing between the approximations of the critical curve as a parabola (cusp) or a line (fold) for a given source position (a) shows that the fold critical point always lies below the cusp critical point in the image plane (d). Depending on the distance of the source to the caustic (a)-(c) and the images to the critical curve (d)-(f), the fold approximation is more precise than the cusp approximation (b) and (e) or vice versa (c) and (f). Image $C$, not being relevant for the argument, is omitted here.}
\label{fig:approximation_comparison}
\end{figure*}

We can also determine the relationship between the approximations of the critical curve of the fold and cusp configuration in the case of a triple-image configuration. As a simple example, we place a source on the $y_1$-axis: $\boldsymbol{y} = (y_s, 0)$. Then, the difference in the $x_2$-coordinate between the critical point of the parabolic approximation $\boldsymbol{x}_\mathrm{c}$ and the critical point of the linear approximation $\boldsymbol{x}_\mathrm{f}$ at the $x_1$-value of the fold critical point between $A$ and $B$ is given by
\begin{equation}
\Delta x_2 = x_{\mathrm{f}2}  \left( \sqrt{1 + \dfrac{\phi_{11}^{(0)} \phi_{2222}^{(0)}} {3\left( \phi_{11}^{(0)}\phi_{2222}^{(0)}-2
 \left(\phi_{122}^{(0)} \right)^2\right) }} -1\right)
 \label{eq:delta_x}
\end{equation}
with 
\begin{equation}
x_{\mathrm{f}2}= \sqrt{\dfrac{-3\phi_{122}^{(0)} y_\mathrm{s}} {2\left( \phi_{11}^{(0)}\phi_{2222}^{(0)}-3 \left(\phi_{122}^{(0)} \right)^2\right) }} \;.
\end{equation}
The derivation of this equation can be found in Appendix~\ref{app:fold_cusp_comparison}. Fig.~\ref{fig:approximation_comparison} (a) and (e) visualise the comparison. Eq.~\eqref{eq:delta_x} implies that the parabolic approximation of the critical curve always has a higher $x_2$-value than the linear approximation and that the difference decreases for decreasing distance to the cusp critical point, i.e. $y_\mathrm{s} \rightarrow 0$. The distance $d$ of the source to the fold $\boldsymbol{y}_\mathrm{f}$ and the cusp singular point $\boldsymbol{y}_\mathrm{c}$ also determines which approximation of the critical curve is the more precise one
\begin{align}
d\left(\boldsymbol{y}, \boldsymbol{y}_\mathrm{f}\right)&\ll d\left(\boldsymbol{y}, \boldsymbol{y}_\mathrm{c}\right) &\rightarrow  \quad &\text{fold configuration} \label{eq:source_comp1} \\
d\left(\boldsymbol{y}, \boldsymbol{y}_\mathrm{c}\right) &\ll d\left(\boldsymbol{y}, \boldsymbol{y}_\mathrm{f}\right) &\rightarrow \quad &\text{cusp configuration} \label{eq:source_comp2}\;.
\end{align}

\noindent
Yet, as the source position is not an observable, we use the lensing equations to find the respective distances in the image plane
\begin{align}
d\left(\boldsymbol{x}_A, \boldsymbol{x}_\mathrm{f}\right)&\ll d\left(\boldsymbol{x}_A, \boldsymbol{x}_\mathrm{c}\right) &\rightarrow  \quad &\text{fold configuration} \label{eq:image_comp1} \\
d\left(\boldsymbol{x}_A, \boldsymbol{x}_\mathrm{c}\right) &\ll d\left(\boldsymbol{x}_A, \boldsymbol{x}_\mathrm{f}\right) &\rightarrow \quad &\text{cusp configuration} \label{eq:image_comp2}\;.
\end{align}
An example for Eqs.~\eqref{eq:source_comp1} and \eqref{eq:image_comp1} is given in (b) and (e) of Fig.~\ref{fig:approximation_comparison} and an example for Eqs.~\eqref{eq:source_comp2} and \eqref{eq:image_comp2} is shown in (c) and (f) of Fig.~\ref{fig:approximation_comparison}. Sect.~\ref{sec:sie_asymmetric_cusp} will treat the case of a three-image configuration in an SIE.

\subsection{Treatment of giant arc(let)s}

Approaching the caustic with the source, the distortions of the images increase and their relative distances decrease. When they merge to a giant arc(let), all equations employing the observables of the individual images become inapplicable, including the time delay. 

In the coordinate system defined in Eq.~\eqref{eq:coordinate_system}, two images enclosing a fold critical point will merge to a straight line with slope $m_\mathrm{f}$ perpendicular to the critical curve, which we denote as arclet. Then, Eq.~\eqref{eq:fold4} is replaced by
\begin{equation}
\dfrac{\phi_{122}^{(0)}}{\phi_{222}^{(0)}} = -\dfrac{1}{m_\mathrm{f}} \;.
\end{equation}
It is the only information we can retrieve from this degenerate fold configuration.

The three  images close to a cusp will merge into a giant arc that can be approximated by the parabola with the slope given by Eq.~\eqref{eq:parabola_slope}. From the fitting procedure, we can also read off the vertex position, obtaining the cusp critical point.

%%%%%%%%%%%%%%%%%%
\section{Simulations}
\label{sec:simulation}

In order to set confidence bounds on the quantities in Eqs.~\eqref{eq:fold1}--\eqref{eq:fold4} and Eqs.~\eqref{eq:cusp1}--\eqref{eq:cusp3}, we simulate a series of gravitational lenses with decreasing degree of symmetry: in this paper, we start with a Navarro-Frenk-White mass density profile (NFW-profile), \cite{bib:NFW}, which is usually chosen as a starting point when modelling lensing dark matter halos at galaxy cluster scale. By this simulation, we analyse the deviations due to the Taylor approximation and the fit of the quadrupole moment to the true image shape for Eqs.~\eqref{eq:fold1}--\eqref{eq:fold4}. Due to the axisymmetry of the model, the images are orthogonal to the critical curve, so that we expect Eqs.~\eqref{eq:fold3} and \eqref{eq:fold4} to be zero. Making the lens heavier than usual galaxy cluster scale lenses, we increase the influence of higher order potential derivatives and we investigate how well the potential derivatives of Eqs.~\eqref{eq:fold1}--\eqref{eq:fold4} are reconstructed from the quadrupole moment of the images.

Subsequently, we simulate a galaxy cluster scale SIE, which is one of the simplest lenses in which cusps occur and in which multiple images are not orthogonal to the critical curve anymore. This lens type is usually employed to model galaxy scale lenses. We use it on the galaxy cluster scale to test the validity of Eqs.~\eqref{eq:fold1}--\eqref{eq:parabola_slope} in the vicinity of folds and cusps where the orientation angles of the images differ from each other. In a subsequent paper, we will use the HERA cluster simulation, \cite{bib:Meneghetti}, as the most realistic test case on galaxy cluster scale currently available. For all multiple images in all simulations to have comparable resolution, we use a redshift of $z_\mathrm{d} = 0.507$ for all lenses, $z_\mathrm{s}=2$ for all sources, and construct the (effective) Einstein radius at which we consider the multiple images to be about $23''$.

\subsection{Simulation of sources}
\label{sec:sources}

We implement a circular Sérsic profile, \cite{bib:Sersic},
\begin{equation}
I(\boldsymbol{y}) = I_\mathrm{e} \, \exp {\left( - b_n \, \left( \left(\frac{\left|\boldsymbol{y} -\boldsymbol{y}_\mathrm{e} \right|}{r_e} \right)^{1/n} -1 \right) \right)}
\label{eq:sersic_profile}
\end{equation}
in the source plane, $\boldsymbol{y}=(y_1, y_2)$, and normalise the intensity to 1 at the peak $\boldsymbol{y}_\mathrm{e}$. We set $n=4$, i.e.\ $b_n = 7.669$, and consider a characteristic radius $r_\mathrm{e} = 25$ px. As source redshift, we set $z_\mathrm{s} = 2$, so that our source model represents simplified elliptical galaxies of some ten kiloparsecs, if we set $1~\mbox{px} = 0.1108''$\footnote{This is the resolution of the HERA cluster, which is employed in all simulations to treat them on equal terms.}, as in the image plane.

\subsection{Simulation of the NFW-lens}
\label{sec:nfw_lens}

We simulate an NFW-profile with $\kappa_\mathrm{s} = 3$ and $r_\mathrm{s} = 33''$ for the normalising convergence and the scale radius in the notation of \cite{bib:Bartelmann2}. As our approach can only describe fold configurations close to radial arcs in axisymmetric models, we set the radial Einstein radius to be $23''$. This implies, that we do not obtain a physically realistic galaxy-cluster-sized gravitational lens with a mass of $2.3 \cdot 10^{15} M_\odot$ inside this radius. Yet, it will clearly show the influence of higher order moments on the extraction of observables and the ratios of potential derivatives.

\begin{figure}[b!]
\centering
  \includegraphics[width=0.45\textwidth]{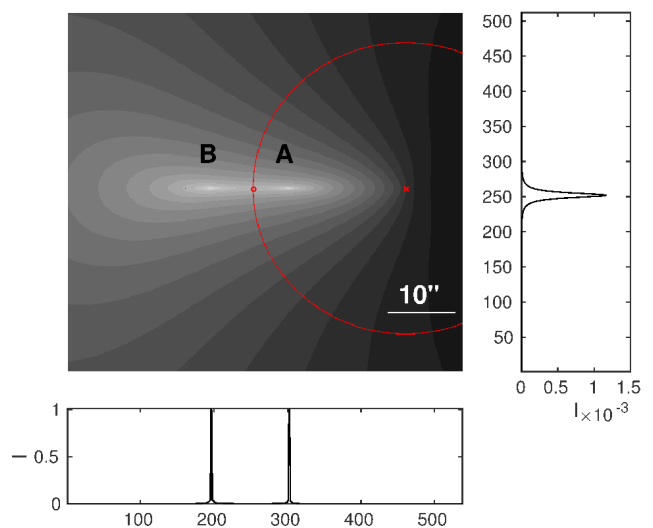}
    \caption{Two-dimensional logarithmic intensity distribution of two images close to the radial critical curve, saddle point image $B$ (left), maximum close to the lens centre $A$ (right), critical curve (red) and centre of the lens (red cross). The plots at the right and bottom show vertical and horizontal intensity profiles through the critical point between $A$ and $B$ (red dot).}
    \label{fig:nfw_example}
\end{figure}

We place 40 sources in steps of 1 px from each other, starting at the caustic going towards the centre of the lens and map them to the image plane, given the lens mapping from the NFW-profile. As result, we obtain two images close to the radial critical curve as shown in Fig.~\ref{fig:nfw_example}, one saddle point image, called image $B$ and one maximum, called image $A$.

\subsection{Simulation of the SIE-lens}
\label{sec:sie_lens}

\begin{figure}[t!]
\centering
  \includegraphics[width=0.45\textwidth]{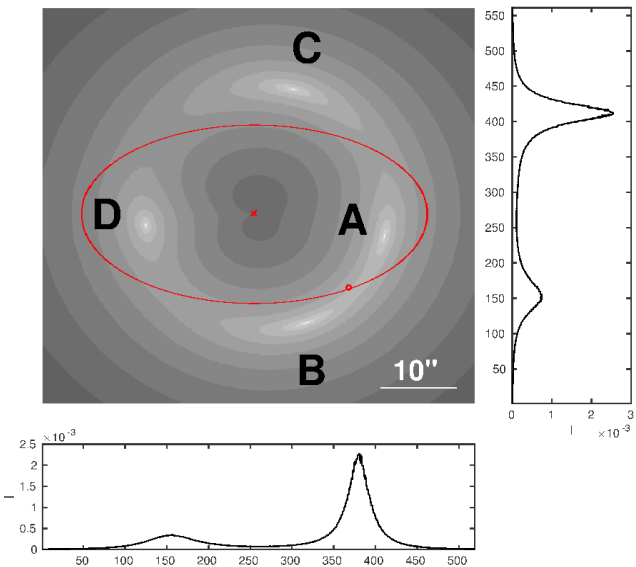}
    \caption{Two-dimensional logarithmic intensity distribution of the four images generated by an SIE. The saddle point image $A$ and the two minima $B$ and $C$ to the right of the lens centre (red cross) are used as cusp configuration, and $A$ and $B$ as configuration close to a fold (red dot). The plots at the right and bottom show vertical and horizontal intensity profiles through the critical point between $A$ and $B$ (red dot).}
    \label{fig:sie_example}
\end{figure}

As second example, we simulate an SIE with an axis ratio $f=0.6091$ and a scale radius of $r_\mathrm{s} = 17''$, as shown in Fig.~\ref{fig:sie_example}. To investigate the Taylor bias at a fold configuration, we choose a fold point between the images $A$ and $B$ (marked by the red dot in Fig.~\ref{fig:sie_example}) and simulate 8 image pairs from Sérsic sources, as described in Sect.~\ref{sec:sources}, with increasing distance from the caustic towards the centre of the lens.

To investigate the cusp configuration of the images $A$, $B$, and $C$, we place Sérsic sources with increasing distance from the cusp in the caustic on the line connecting the lens centre and the cusp.

In addition, we simulate two configurations close to the cusp in which the source is at the same $y_1$, but $y_2$ is once chosen to be zero and once off the symmetry axis, so that $A$ and $B$ approach each other to test the statements made in Sect.~\ref{sec:cusp_fold}.

\subsection{Extraction of observables}
\label{sec:extraction_of_observables}

In order to determine the observables from the multiple images close to the critical curve, we employ SExtractor, \cite{bib:Bertin}. We use the centre of light of the images as relative distance of the images, and determine their ellipticities and orientation angles from their second order moments. In order to calculate them SExtractor requires a segmentation into foreground objects (gravitationally lensed images) and background. As our data is noise free, we only threshold the raw images without applying a median filter or a background subtraction. The observables determined from second order moments depend on the threshold and we will investigate the influence of the threshold on the accuracy of the reconstruction in Sect.~\ref{sec:nfw_fold}. 

The observables for all images are subsequently transformed into the coordinate system defined by Eq.~\eqref{eq:coordinate_system} and then inserted into Eqs.~\eqref{eq:fold1}--\eqref{eq:parabola_slope}.

\subsection{Evaluation}
\label{sec:evaluation}

\subsubsection{Measures for the accuracy of the approach}
\label{sec:evaluation_criteria}

To estimate the accuracy of the time delay we replace
\begin{equation}
c t_{AB}/\Gamma_\mathrm{d} = \Delta \phi \equiv \phi(\boldsymbol{x}_A,\boldsymbol{y}) -\phi(\boldsymbol{x}_B,\boldsymbol{y}) 
\label{eq:time_delay_replacement}
\end{equation}
in Eqs.~\eqref{eq:fold1} and \eqref{eq:cusp1} and calculate the potential difference between two images $A$ and $B$ from the simulated model.

As measure for the deviation, we divide the (ratios of) potential derivatives as obtained from Eqs.~\eqref{eq:fold1}--\eqref{eq:fold4} and Eqs.~\eqref{eq:cusp1}--\eqref{eq:cusp3} by their true values given by the lens model. Analogously, we assess the deviation between Eq.~\ref{eq:parabola_slope} and the respective ratio of true ratios of potential derivatives in the cusp case. Deviations of the parabolic approximation from the critical curve are shown in graphs in Sects.~\ref{sec:sie_cusp} and \ref{sec:sie_asymmetric_cusp}.

The difference between the true fold position, $\boldsymbol{x}_\mathrm{t}$, and the reconstructed one, $\boldsymbol{x}_\mathrm{r}$, is calculated as 
\begin{equation}
\Delta \boldsymbol{x}_{0} \equiv \boldsymbol{x}_\mathrm{t} - \boldsymbol{x}_\mathrm{r} = \boldsymbol{x}_\mathrm{t} - \dfrac{\delta_{AB2}}{2} \;,
\label{eq:x_0_reconstruction}
\end{equation}
using the distance between the intensity mean of $A$ and $B$ as $\delta_{AB2}$. To determine the cusp critical point we employ Eqs.~\eqref{eq:cusp_coords1} and \eqref{eq:cusp_coords2} and subtract the results from the position of image $A$.

\subsubsection{Evaluation NFW (fold configuration)}
\label{sec:nfw_fold}

\begin{figure*}[ht]
\centering
\begin{subfigure}{0.32\textwidth}
  \centering
   \includegraphics[width=0.95\linewidth]{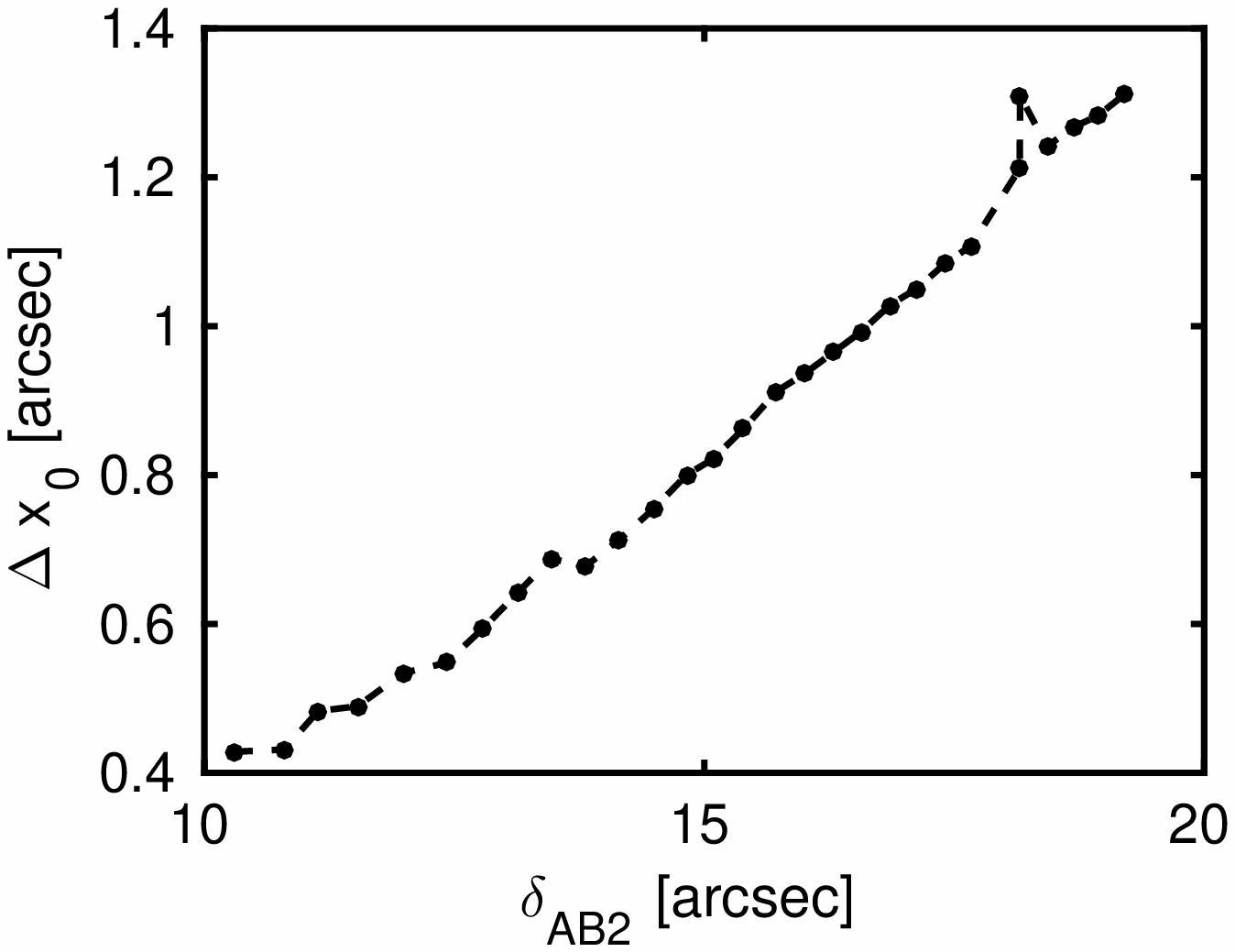}
  \caption{}
  \end{subfigure}
  \begin{subfigure}{0.32\textwidth}
  \centering
   \includegraphics[width=0.95\linewidth]{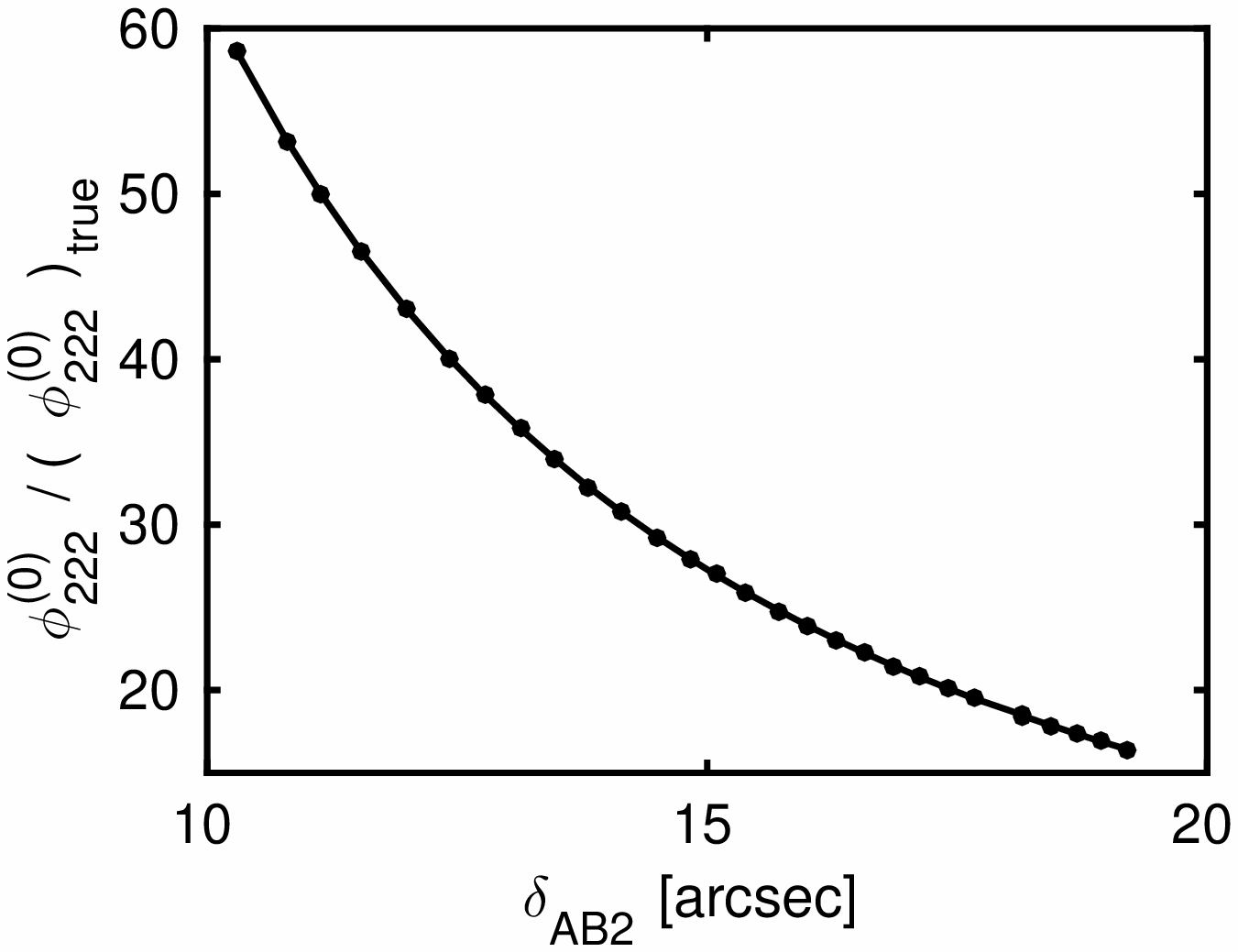}
  \caption{}
  \end{subfigure}
    \caption{Evaluation of the NFW-profile: Deviation of the reconstructed critical point from the true one, as defined in Eq.~\eqref{eq:x_0_reconstruction}, dependent on the (observable) relative distance between the intensity mean of the two images (a). Accuracy of Eq.~\eqref{eq:fold1} dependent on the same distance between images $A$ and $B$ (b).}
 \label{fig:nfw1}
\end{figure*}

\begin{figure*}[h!]
\centering
\begin{subfigure}{0.325\textwidth}
  \centering
  \includegraphics[width=\linewidth]{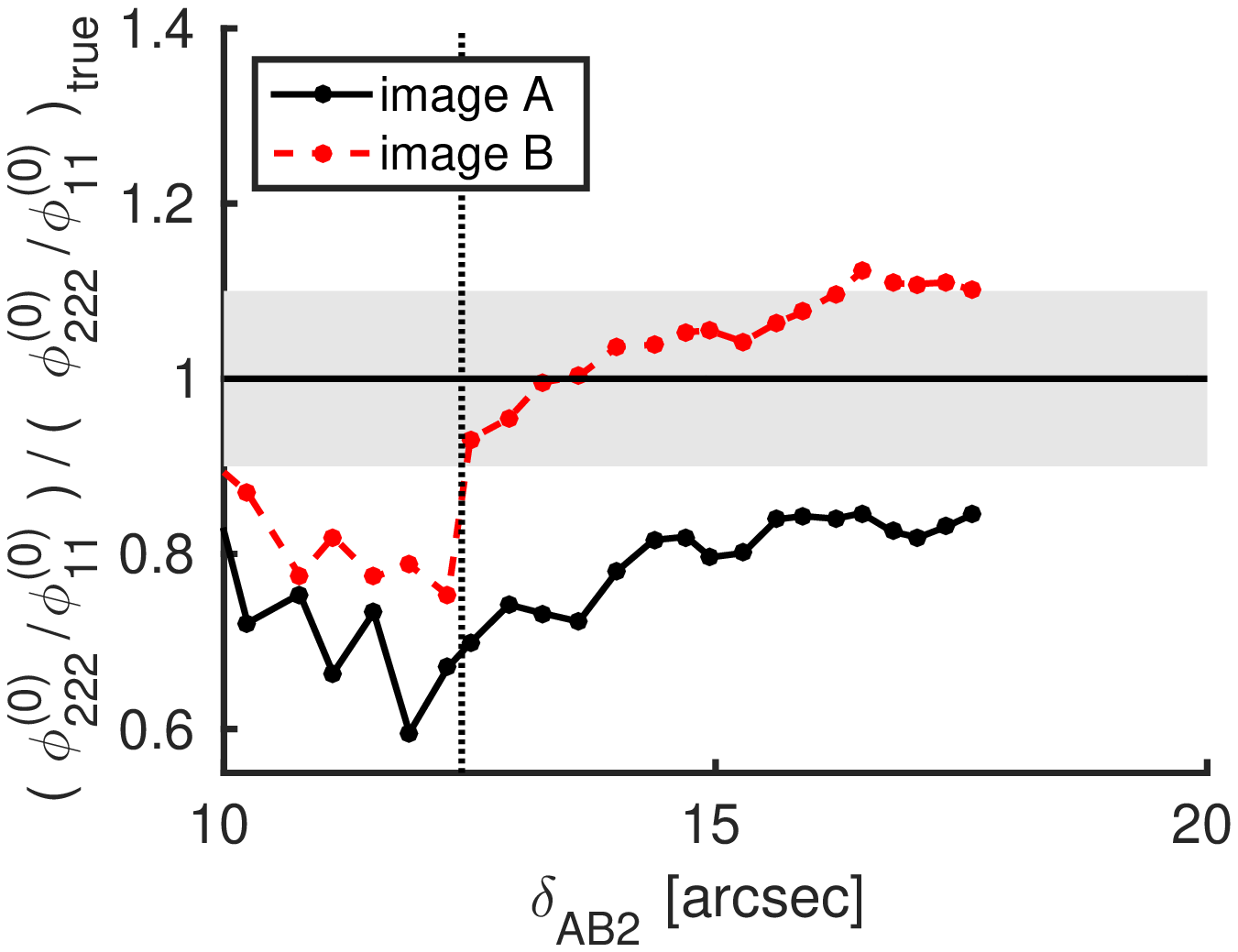}
  \caption{}
  \end{subfigure}
  \begin{subfigure}{0.325\textwidth}
  \centering
  \includegraphics[width=\linewidth]{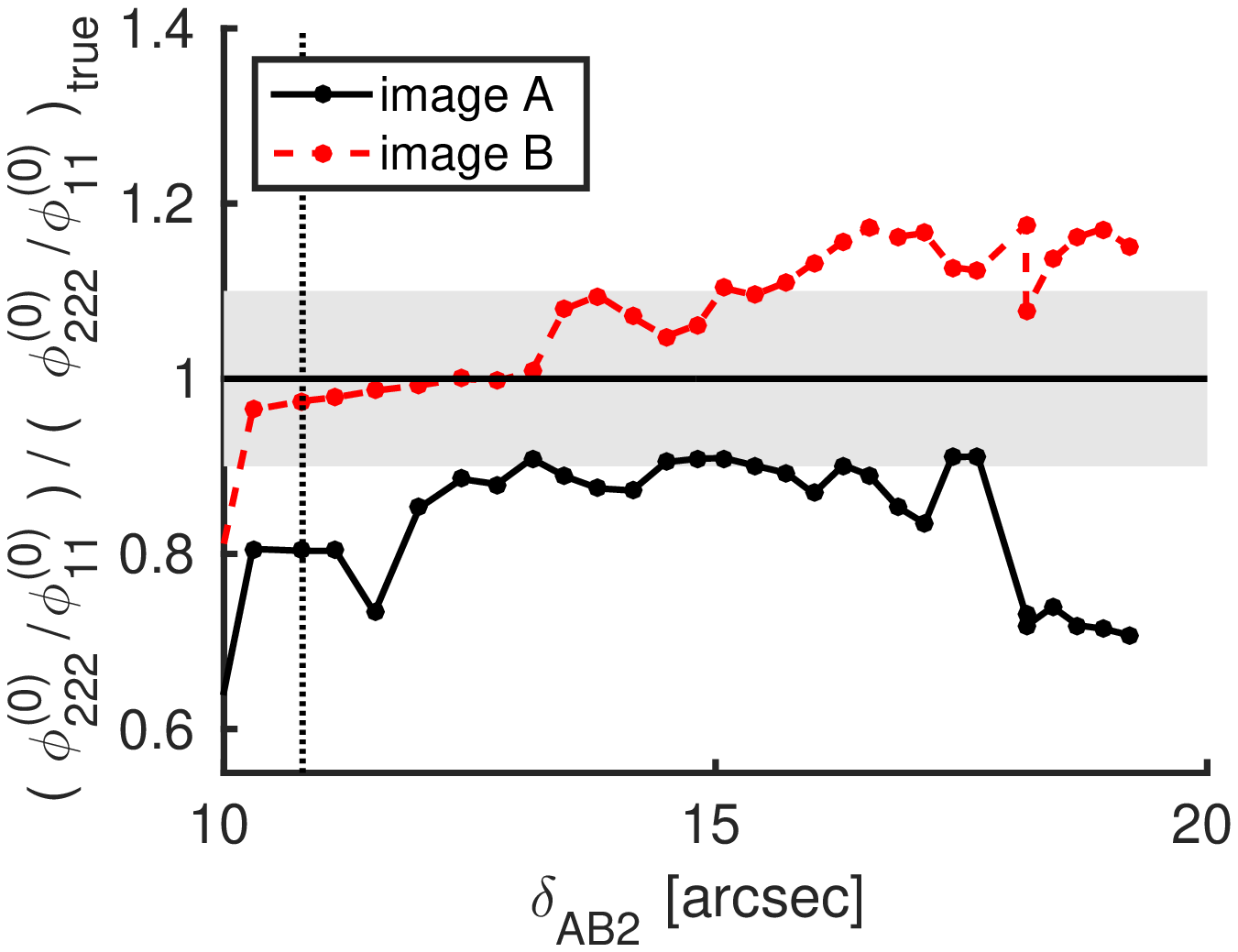}
  \caption{}
  \end{subfigure}
  \begin{subfigure}{0.325\textwidth}
  \centering
  \includegraphics[width=\linewidth]{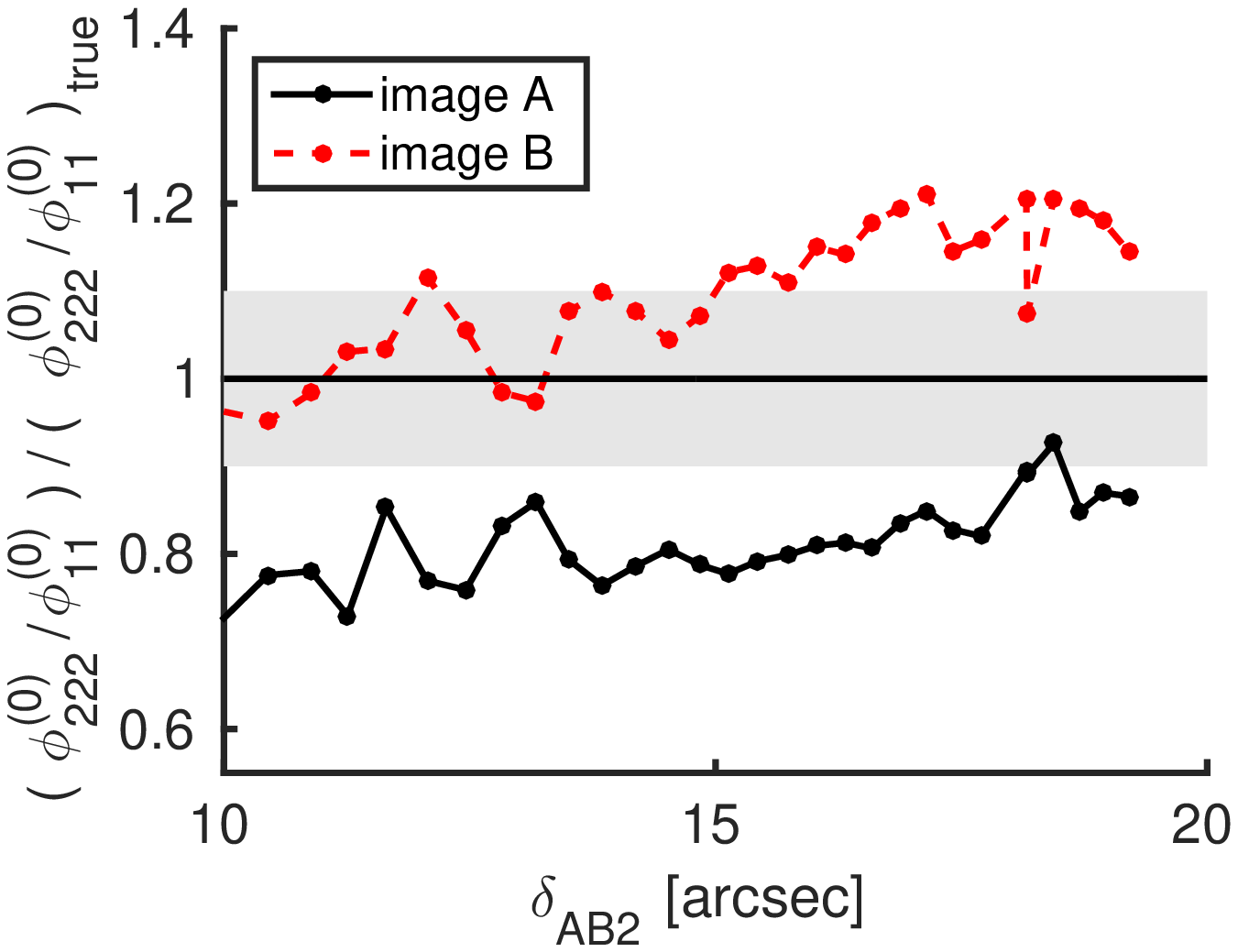}
  \caption{}
  \end{subfigure}
    \caption{Evaluation of the NFW-profile: Ratio of Eq.~\eqref{eq:fold2} divided by the true value for image $A$ (black, solid line) and image $B$ (red, dashed line) for a low SExtractor threshold (a), an intermediate threshold (b), and a high threshold (c). Vertical dotted lines indicate the minimum distance at which $A$ and $B$ are thresholded as two separate objects. The grey-shaded area marks 10\% deviation from the true ratio.}
    \label{fig:phi222_phi11_accuracy}
\end{figure*}

First, we evaluate the accuracy of the reconstruction of the critical point, as shown in Fig.~\ref{fig:nfw1}. 
From Fig.~\ref{fig:nfw1} (a), we read off that the reconstructed critical point is systematically shifted towards the saddle point image (as the positive $x_1$-axis points from $B$ to $A$). The deviation increases for increasing distance of the images from the critical curve, i.e.\ the expansion point of the Taylor series of the lensing potential. The reconstructed position of the critical point is biased towards image $B$ because its distance to the critical point increases faster than the distance of image $A$ to the critical point when moving the source away from the caustic.

Next, we evaluate the accuracy of Eq.~\eqref{eq:fold1} using Eq.~\eqref{eq:time_delay_replacement} and plot the results in Fig.~\ref{fig:nfw1} (b). The accuracy decreases with decreasing distance from the critical curve as $\delta_{AB2}$ in the denominator of Eq.~\eqref{eq:fold1} approaches zero. The steep decrease of accuracy is consistent with the results for point sources in elliptical models, as shown in \cite{bib:Wagner}, in which we found that the deviation between $1/12 \, \delta_{AB2}^3 \phi_{222}^{(0)}$ from the potential difference $\Delta \phi$ increases for increasingly axisymmetric lenses.

Then, we insert the observables into Eq.~\eqref{eq:fold2} to determine its accuracy. Fig.~\ref{fig:phi222_phi11_accuracy} shows the results for increasing threshold $5 \cdot 10^4,\, 2 \cdot 10^5,\, 3 \cdot 10^5$ (a) to (c). For small relative image separations (marked by the vertical dotted lines in Fig.~\ref{fig:phi222_phi11_accuracy}), the threshold is too low, so that SExtractor calculates the second order moments from two merging images, so that our approach cannot be applied due to incorrectly determined observables. Between relative image separations of $10''$ and $20''$ lies the range of applicability of our approach. Farther away from the critical curve, the assumption $|\mu_A| = |\mu_B|$ breaks down within the precision set by an HST-like gain of 2 to determine the uncertainty of the image fluxes in the SExtractor configuration file. 

Comparing the results for image $A$ and $B$, we notice that the ratio of potential derivatives is more accurate for image $B$ for any threshold because $A$ is smaller and at the resolution limit to extract $r_A$. For increasing distances from the critical curve, the images also become fainter so that the number of pixels belonging to one image, i.e.\ its resolution, is decreased for a fixed threshold. The alternating local peaks and minima in the curves for $A$ are effects of the pixelisation.  The same effect is clearly observed for $B$ for the highest threshold and mildly at large distances for the intermediate threshold. 
Focussing on $B$, the true ratio of potential derivatives can be reconstructed with deviations less than 22\% for all thresholds. As $\delta_{AB2}$ is the same for all thresholds and the isocontours of the images are self-similar for fixed $\delta_{AB2}$, the differences between the graphs in Fig.~\ref{fig:phi222_phi11_accuracy} show the influence of the pixelisation and resolution of $r$. 

Due to the unphysically high mass of the simulated NFW-lens, the distorted images show features of flexion that will not be as pronounced in realistic images. Hence, retrieving the ratios of potential derivatives shown in Fig.~\ref{fig:phi222_phi11_accuracy} to such a high accuracy shows that possible influences of higher order shape distortions to the quadrupole moment cannot be large.

As images $A$ and $B$ are oriented orthogonal to the critical curve, Eq.~\eqref{eq:fold3} always yields zero, resulting in a slope of zero (Eq.~\eqref{eq:fold4}). We omit the graphs for these ratios of potential derivatives here, as the orthogonality is reproduced well because SExtractor correctly determines $|\varphi_i| = \pi/2,\, i=A, B$ in all cases.

\subsubsection{Evaluation SIE (fold configuration)}
\label{sec:sie_fold}

\begin{figure*}[ht]
\centering
\begin{subfigure}{0.32\textwidth}
  \centering
   \includegraphics[width=0.95\linewidth]{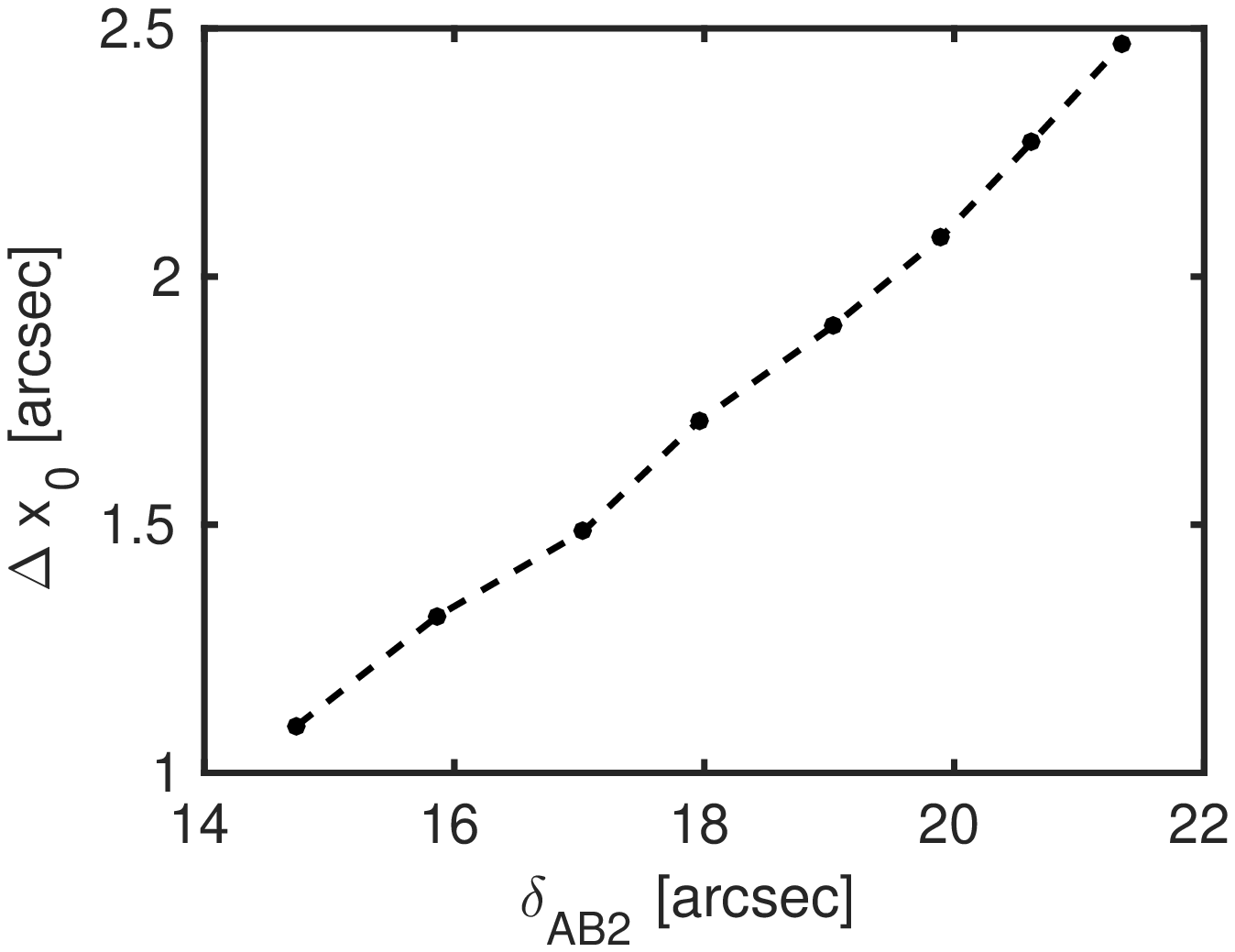}
  \caption{}
  \end{subfigure}
  \begin{subfigure}{0.32\textwidth}
  \centering
   \includegraphics[width=0.95\linewidth]{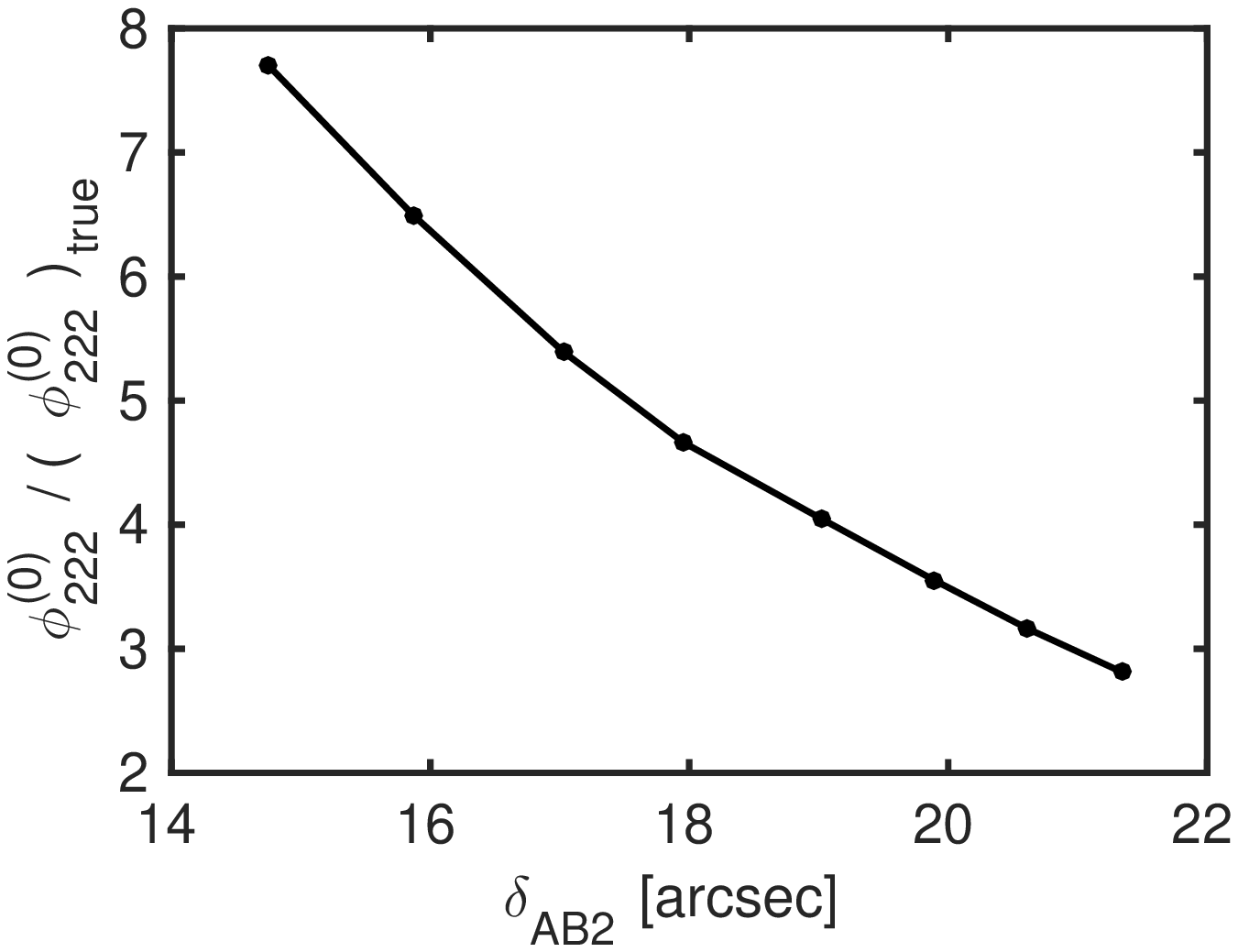}
  \caption{}
  \end{subfigure}
    \caption{Evaluation of the SIE-profile (fold): Deviation of the reconstructed critical point from the true one, as defined in Eq.~\eqref{eq:x_0_reconstruction}, dependent on the (observable) relative distance between the intensity mean of the two images (a). Accuracy of Eq.~\eqref{eq:fold1} dependent on the same distance between images $A$ and $B$ (b).}
 \label{fig:sie1}
\end{figure*}

\begin{figure*}[h!]
\centering
\begin{subfigure}{0.325\textwidth}
  \centering
  \includegraphics[width=\linewidth]{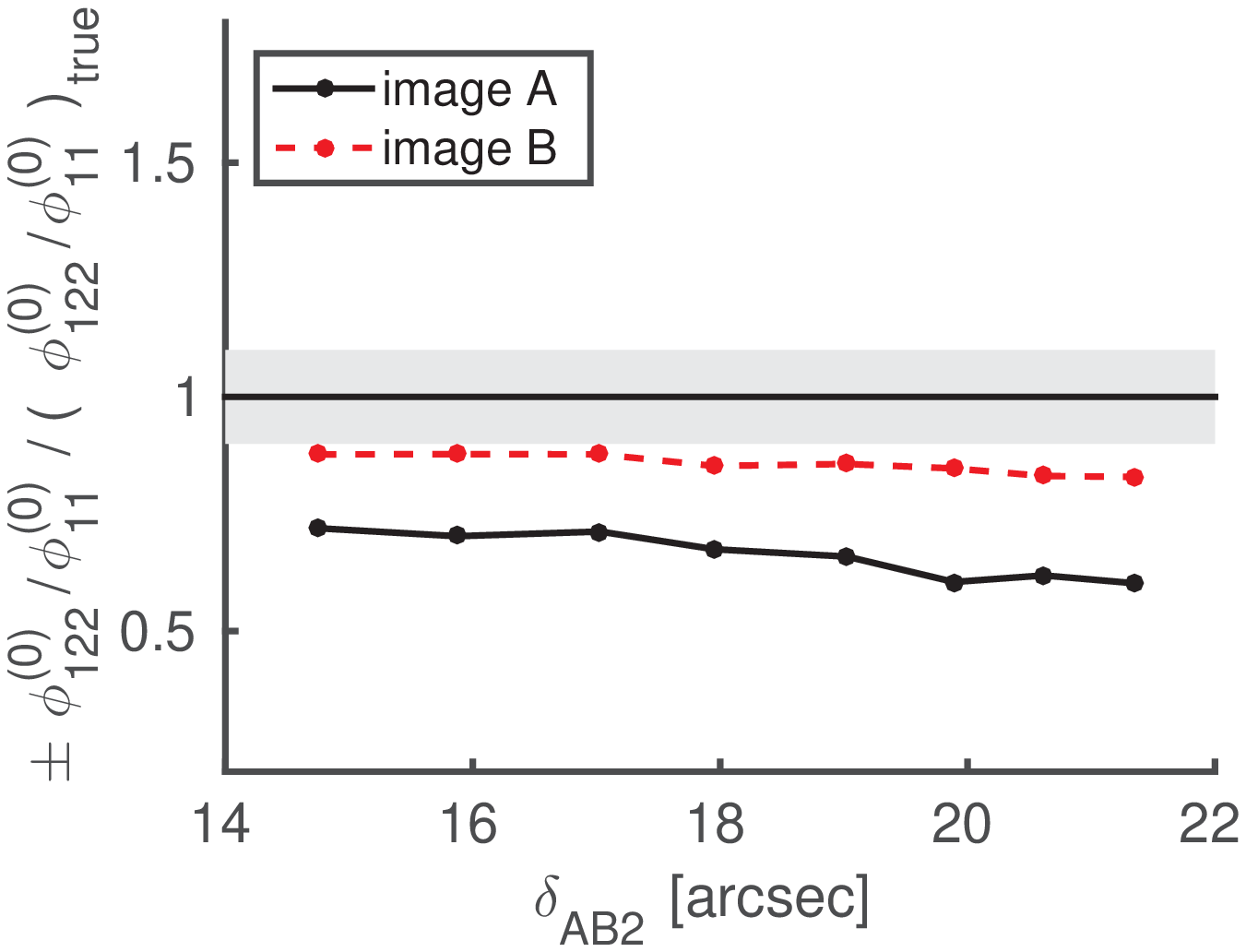}
  \caption{}
  \end{subfigure}
  \begin{subfigure}{0.325\textwidth}
  \centering
  \includegraphics[width=\linewidth]{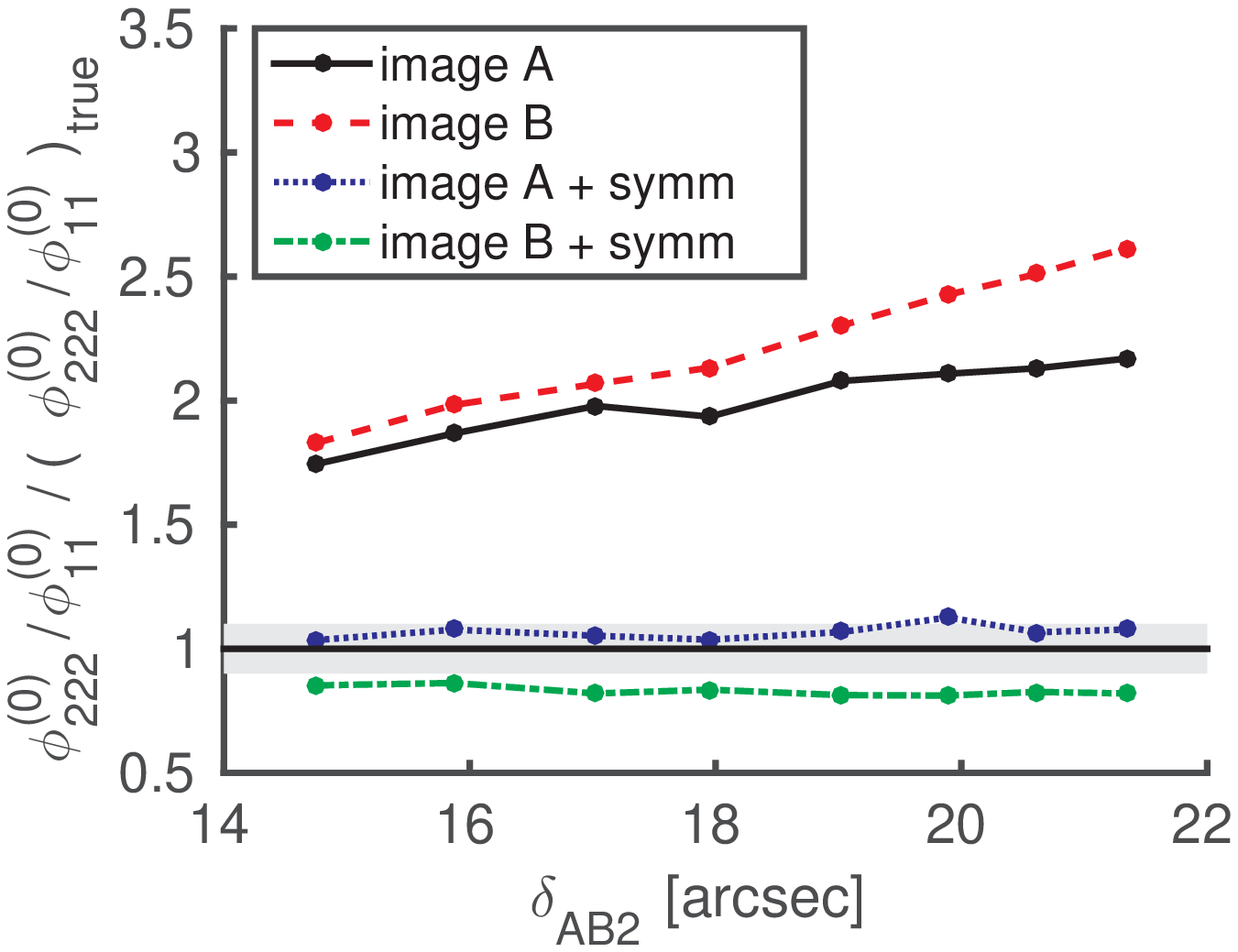}
  \caption{}
  \end{subfigure}
  \begin{subfigure}{0.325\textwidth}
  \centering
  \includegraphics[width=\linewidth]{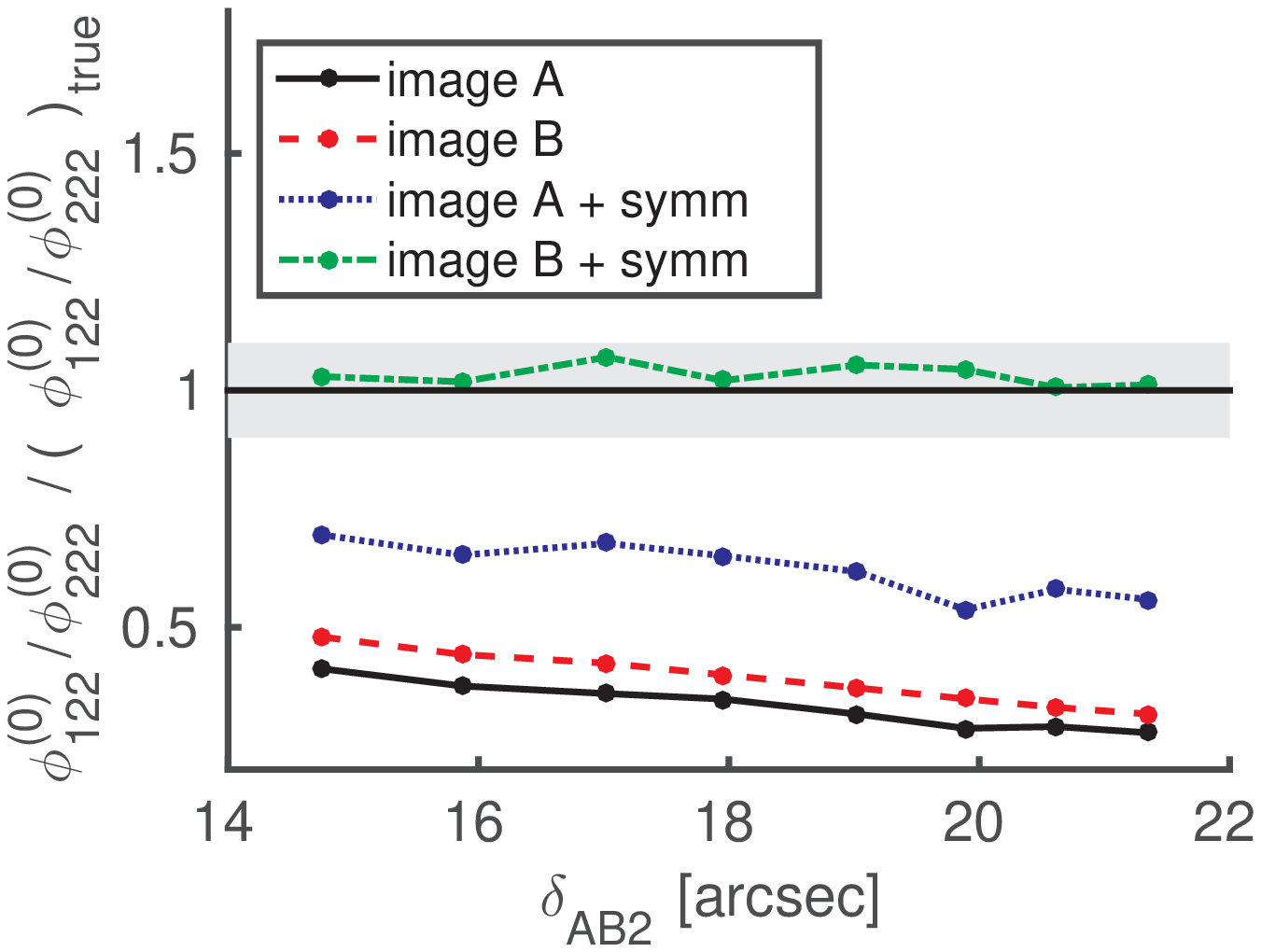}
  \caption{}
  \end{subfigure}
    \caption{Evaluation of the SIE-profile (fold): Ratio of Eq.~\eqref{eq:fold3} divided by the true value for image $A$ (black, solid line) and image $B$ (red, dashed line) (a), ratio of Eq.~\eqref{eq:fold2} for image $A$ and $B$ and for both images under the symmetry assumption that their orientation is orthogonal to the critical curve (blue dotted line, green dash-dotted line) (b), slope of the critical curve, Eq.~\eqref{eq:fold4}, at the fold using the ratios obtained in (a) and (b) (c). The grey-shaded area marks 10\% deviation from the true ratio.}
    \label{fig:sie_ratios}
\end{figure*}

Analogously, we proceed for the SIE, determine one SExtractor threshold for all simulated multiple images at increasing distance from the fold such that pixelisation effects are negligible and extract the observables. Inserting them into Eqs.~\eqref{eq:fold1}--\eqref{eq:fold4}, we obtain the results shown in Figs.~\ref{fig:sie1} and \ref{fig:sie_ratios}. Comparing to the (unrealistic) NFW-profile, the inaccuracy of reconstruction of the critical curve is increased by about a factor of two. The inaccuracy of $\phi_{222}^{(0)}$ is decreased by roughly a factor of ten, but still too high to be of use to break the mass-sheet-degeneracy. Comparing this accuracy to the one for $\phi_{11}^{(0)}$ and $\phi_{2222}^{(0)}$ achievable close to cusps in Sect.~\ref{sec:sie_cusp}, we note that the expansion of the potential to third order at folds is no sufficient approximation and higher orders have to be included.

From Fig.~\ref{fig:sie_ratios} (a), we read off that the observables from the image with positive parity, $B$, approximate $\phi_{122}^{(0)}/\phi_{11}^{(0)}$ best with deviations between 13\% and 18\% from the true value. (b) shows that the assumption of images orthogonally oriented to the critical curve yields accuracies of $\phi_{222}^{(0)}/\phi_{11}^{(0)}$ in the range of 90\% and including information about the orientation angle yields worse results. Consequently, the accuracy of the slope of the critical curve at the fold is highest for image $B$ using the orthogonality assumption to determine $\phi_{222}^{(0)}/\phi_{11}^{(0)}$. In all three graphs, the accuracy of the reconstruction using the image orientation angles visibly decreases with increasing distance from the fold point where other critical points gain influence. Hence, the image orientation angles and their difference give hints at local assymmetries in the lens map. Using the symmetry assumption of images orientated orthogonal to the critical curve for Eq.~\eqref{eq:fold2}, its accuracy is almost constant over the distance range, supporting this statement.

On the whole, the highest accuracy for all properties of the critical curve that can be retrieved by our approach at a fold point is obtained by using the orientation angle of the images only when calculating Eq.~\eqref{eq:fold3} and assuming $\varphi_i = \pi/2$, $i=A,B$, in Eq.~\eqref{eq:fold2}.

\subsubsection{Evaluation SIE (symmetric cusp configuration)}
\label{sec:sie_cusp}

Next, we investigate the accuracy of Eqs.~\eqref{eq:cusp1}--\eqref{eq:parabola_slope} for three-image-configurations generated by sources placed at increasing distance on the symmetry axis connecting the lens centre and the cusp in the caustic, as detailed in Sect.~\ref{sec:sie_lens}. We also compare the accuracy achievable by  Eqs.~\eqref{eq:cusp1}--\eqref{eq:parabola_slope} with the accuracy that is obtained by the approach stated in \cite{bib:Wagner} which makes the symmetry assumption that the semi-major axes of all images are oriented parallel to the critical curve at the cusp. This assumption amounts to the Taylor-expanded lens mapping detailed in \cite{bib:Petters} and it's range of validity around the cusp point is smaller than the range of validity of the approach described in \cite{bib:SEF}, on which our current approach is based.

\begin{figure*}[ht]
\centering
\begin{subfigure}{0.32\textwidth}
  \centering
   \includegraphics[width=0.95\linewidth]{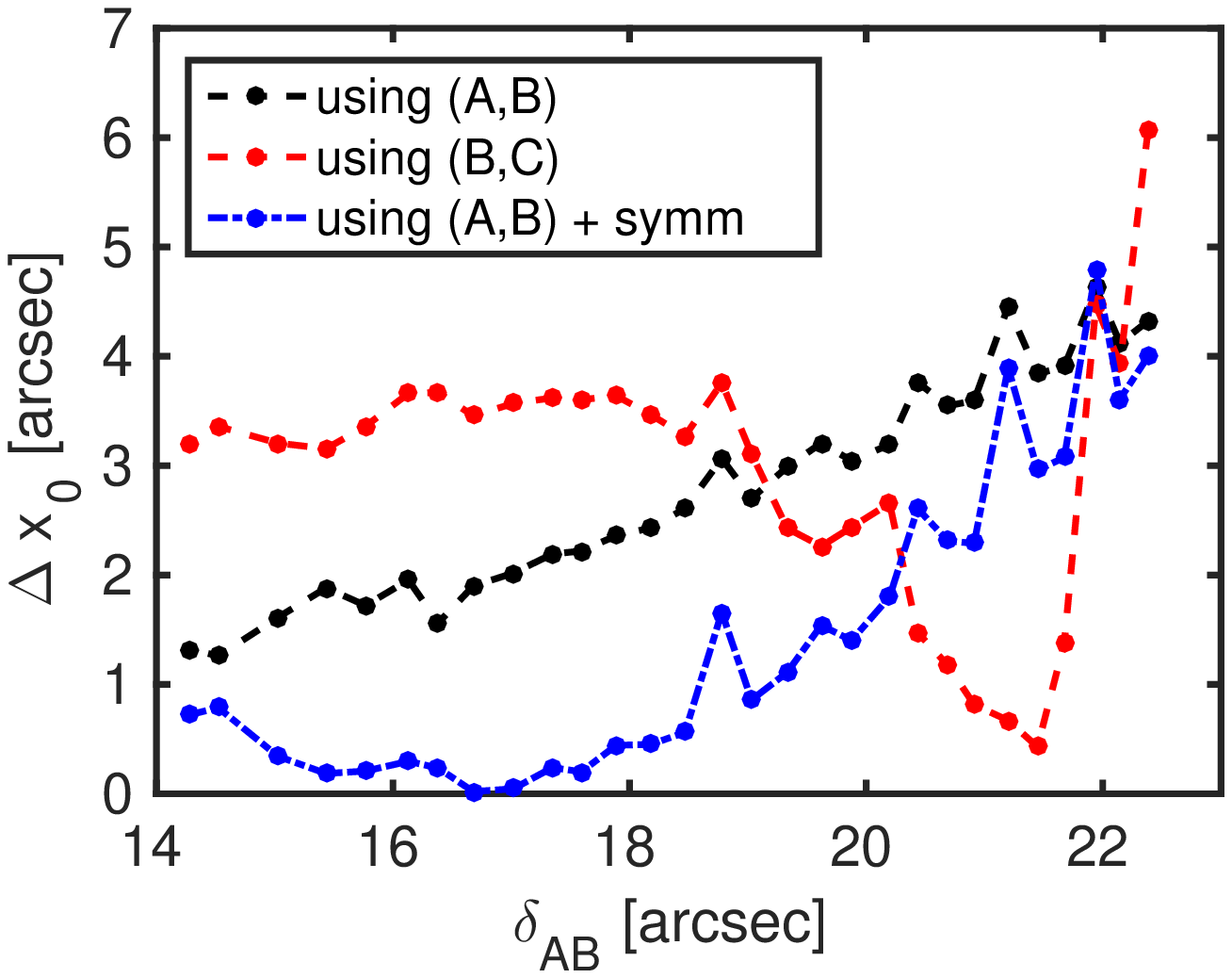}
  \caption{}
  \end{subfigure}
  \begin{subfigure}{0.32\textwidth}
  \centering
   \includegraphics[width=0.95\linewidth]{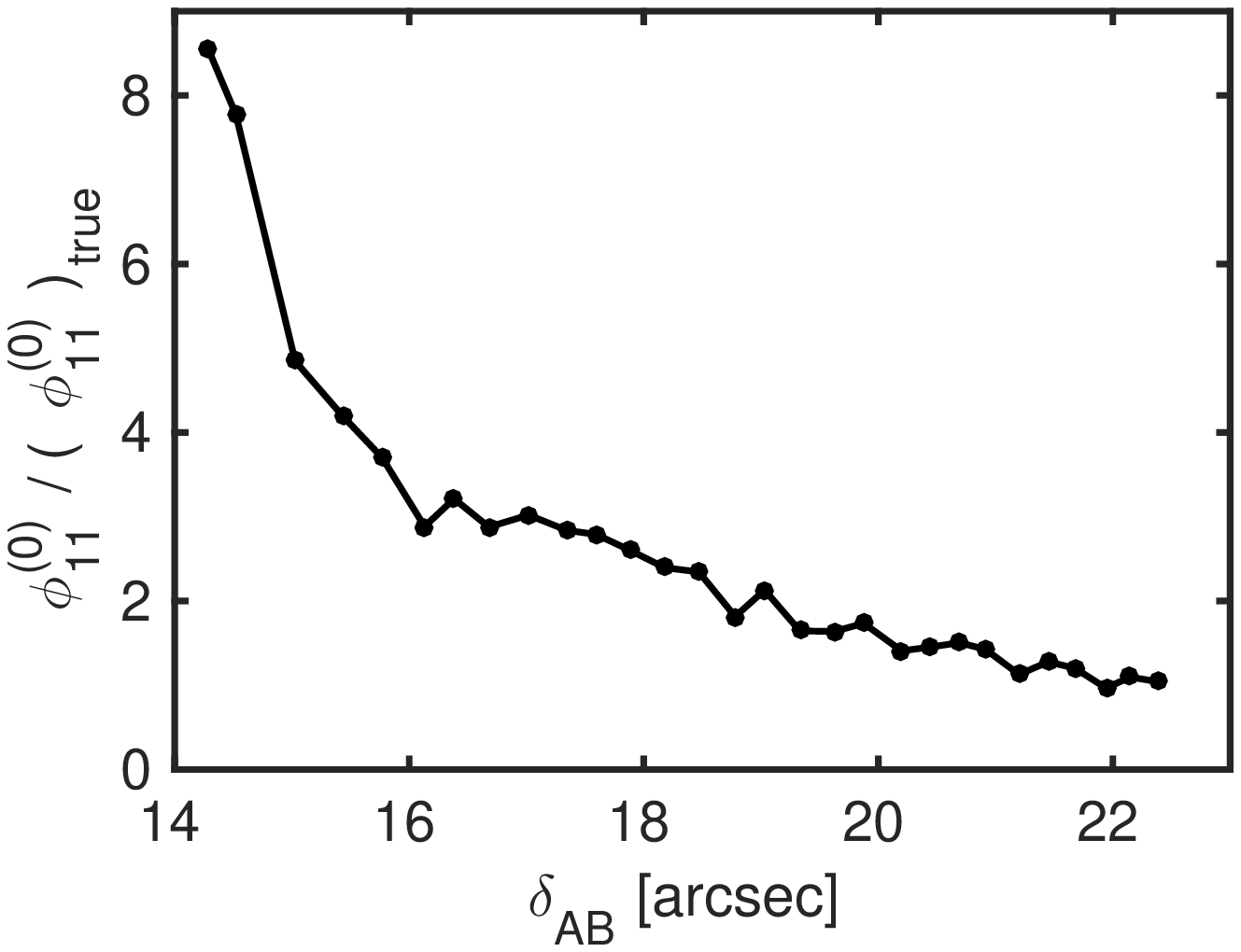}
  \caption{}
  \end{subfigure}
  \begin{subfigure}{0.32\textwidth}
  \centering
   \includegraphics[width=0.95\linewidth]{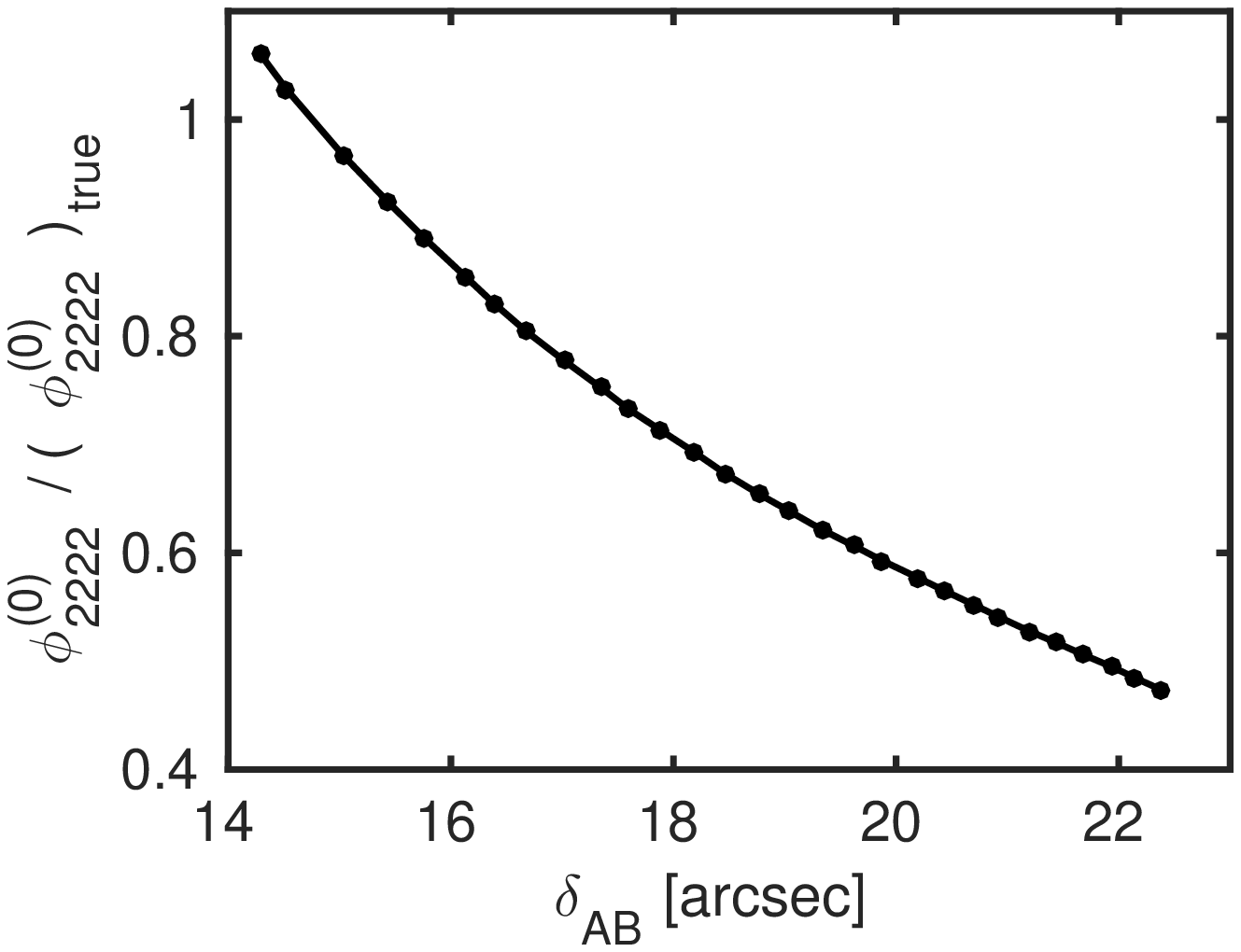}
  \caption{}
  \end{subfigure}
    \caption{Evaluation of the SIE-profile (cusp): Deviation of the reconstructed critical point from the true one dependent on the (observable) relative distance between the intensity mean of the two images $A$ and $B$ using the pair $(A,B)$ (black, solid line), the pair of positive parity images $(B,C)$ (red, dashed line) or the pair $(A,B)$ with the symmetry assumption that image orientation angles are parallel to the critical curve at the cusp (blue, dash-dotted line) (a). Using $(A,C)$ yields the same result as $(A,B)$ because the source lies on the symmetry axis between $B$ and $C$. Accuracy of Eq.~\eqref{eq:cusp1} dependent on the same distance between images $A$ and $B$ (b) and accuracy of the analogue to Eq.~\eqref{eq:cusp1} under the symmetry assumption as stated in \cite{bib:Wagner} (c).}
 \label{fig:sie2}
\end{figure*}

\begin{figure*}[h!]
\centering
\begin{subfigure}{0.325\textwidth}
  \centering
  \includegraphics[width=\linewidth]{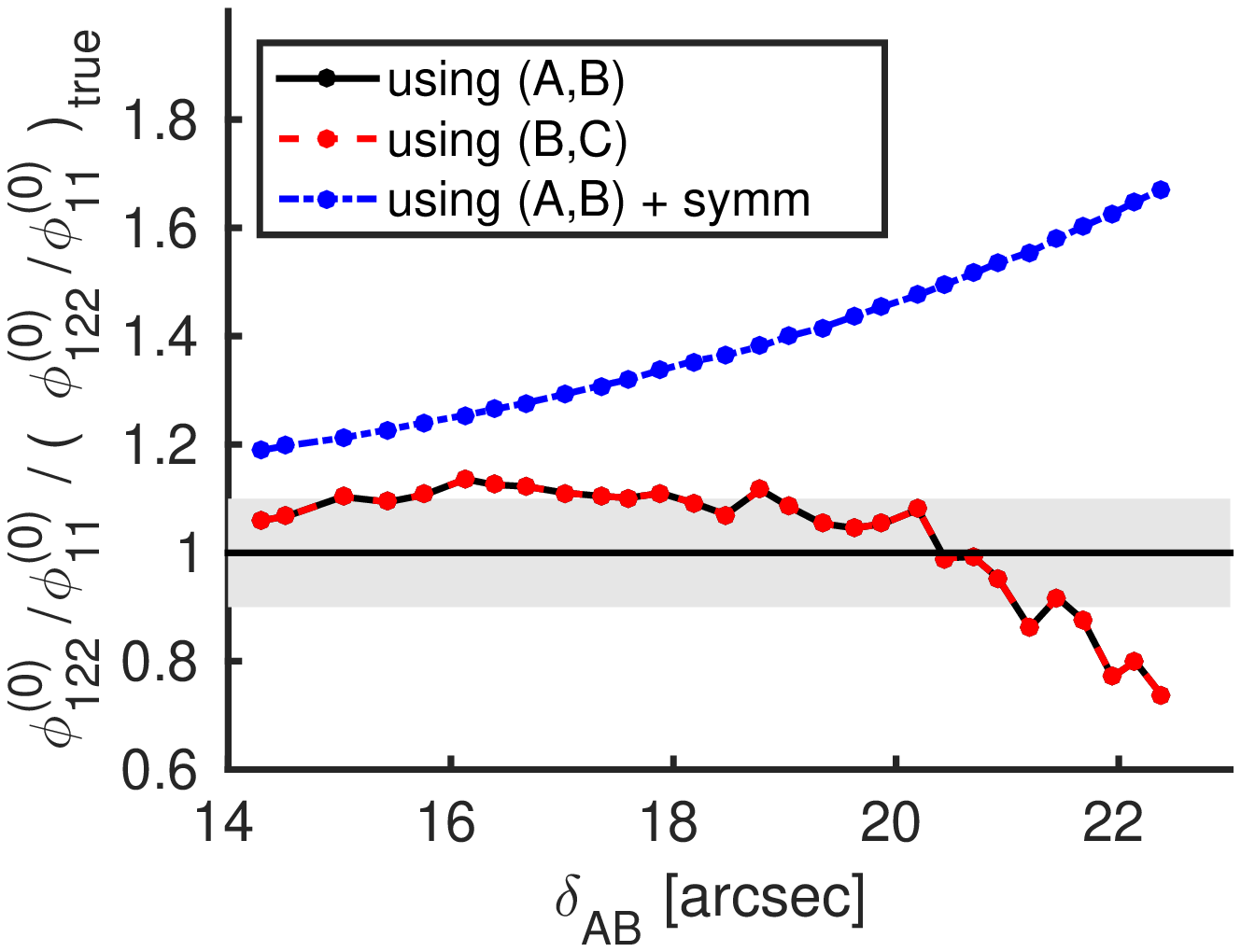}
  \caption{}
  \end{subfigure}
  \begin{subfigure}{0.325\textwidth}
  \centering
  \includegraphics[width=\linewidth]{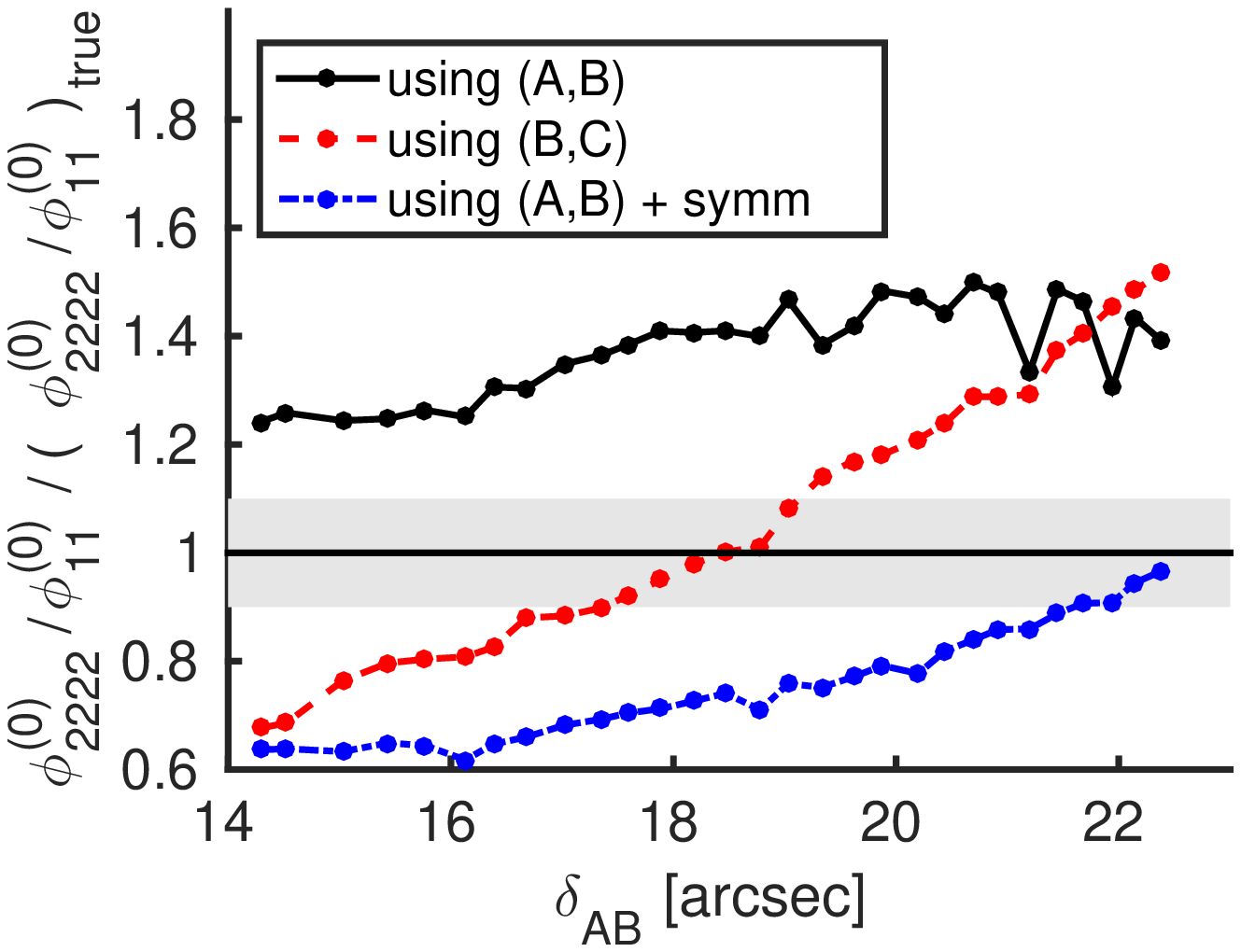}
  \caption{}
  \end{subfigure}
  \begin{subfigure}{0.325\textwidth}
  \centering
  \includegraphics[width=\linewidth]{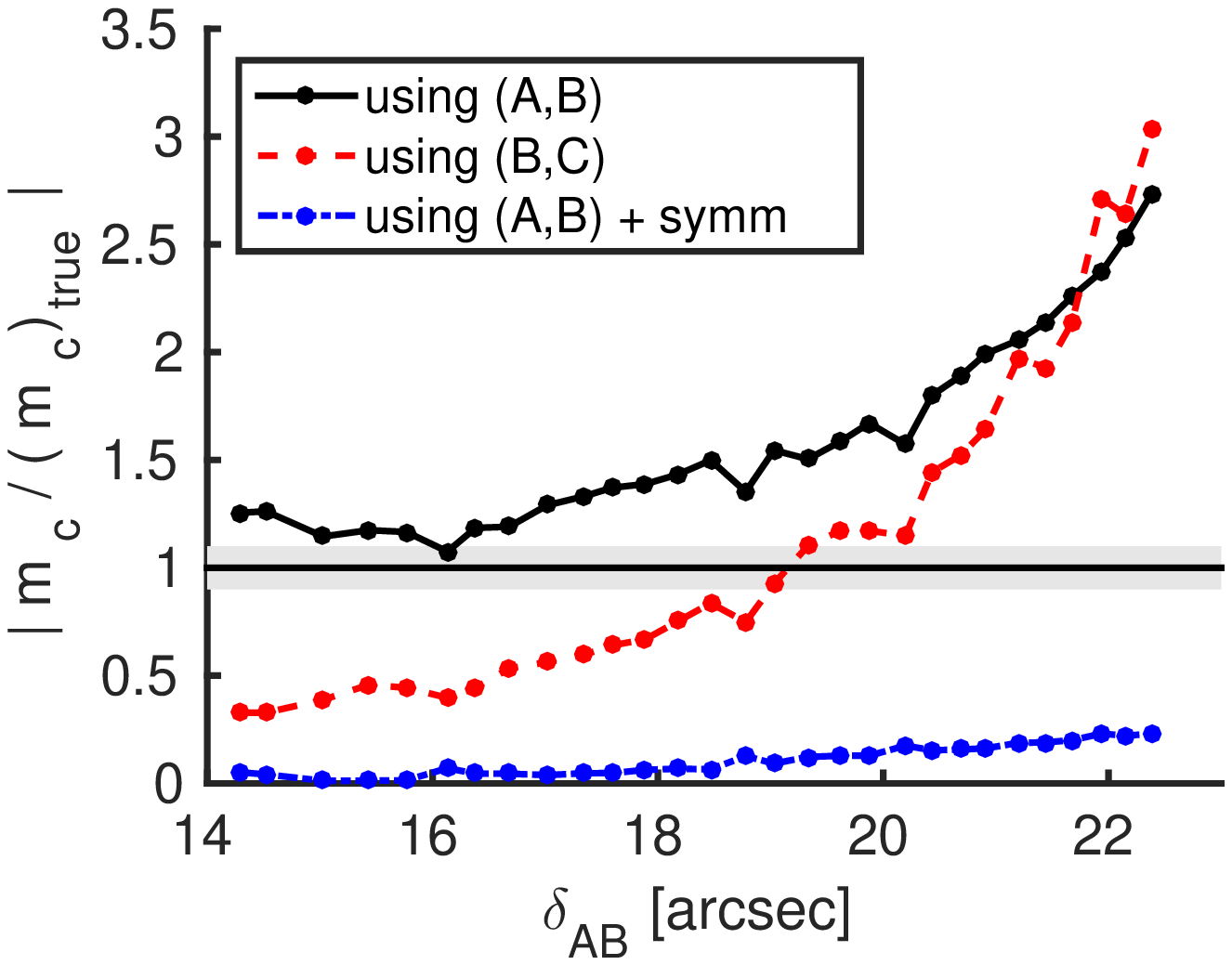}
  \caption{}
  \end{subfigure}
    \caption{Evaluation of the SIE-profile (cusp): Ratio of Eq.~\eqref{eq:cusp2} divided by the true value for the same configurations of image pairs and the symmetry assumption as in Fig.~\ref{fig:sie2}, the result for $(A,B)$ (black, solid line) coincides with the one for $(B,C)$ (red, dashed line) (a), ratio of Eq.~\eqref{eq:cusp3} (b), slope of the critical curve at the cusp, Eq.~\eqref{eq:parabola_slope}, using the ratios obtained in (a) and (b) (c). The grey-shaded area marks 10\% deviation from the true ratio.}
    \label{fig:sie_ratios2}
\end{figure*}

Fig.~\ref{fig:sie2} (a) shows the accuracy of the cusp point reconstruction in the image plane inferred from different combinations of image pairs. As expected, using the two positive parity images $B$ and $C$ to reconstruct the cusp position is the most unreliable estimate, while the approach stated in \cite{bib:Wagner} using the pair $(A, B)$ excels over the approach including the orientation angles for the same image pair close to the critical curve. This result can also be seen in Fig.~\ref{fig:asymmetric_cusp} (a) comparing the position of the vertices of the parabolic approximations to the critical curve. Yet, as (b) shows, taking the orientation angles of the images into account yields a more accurate estimate of the cusp point when the symmetry of the image configuration is broken, i.e.\ for sources that do not lie on the axis connecting the lens centre and the cusp in the caustic.

As Eq.~\eqref{eq:cusp1} diverges at the critical curve, like Eq.~\eqref{eq:fold1}, the inaccuracy is of the same order as that of Eq.~\eqref{eq:fold1}. Fig.~\ref{fig:sie2} (b) and (c) shows that breaking the mass-sheet-degeneracy may only be achieved to an accuracy below 10\% in a small interval of relative distances between $A$ and $B$ under the symmetry assumption stated in \cite{bib:Wagner}. However, as Fig.~\ref{fig:sie_ratios2} shows, the accuracies of Eqs.~\eqref{eq:cusp2}, \eqref{eq:cusp3}, and \eqref{eq:parabola_slope} are on the order of 5\% to 25\%, such that the resulting potential derivatives (convergence, shear, and flexion) are still prone to high inaccuracies.

All graphs in Fig.~\ref{fig:sie_ratios2} clearly show that the accuracy of the ratios of potential derivatives is vastly improved, if image orientation angles are taken into account. This can be visually confirmed in Fig.~\ref{fig:asymmetric_cusp} comparing the reconstruction of the critical curve according to \cite{bib:Wagner} (green curve) with the reconstructions obtained from different combinations of images including their orientation angles.

\subsubsection{Evaluation SIE (asymmetric cusp configuration)}
\label{sec:sie_asymmetric_cusp}

\begin{figure*}[ht]
\centering
\begin{subfigure}{0.495\textwidth}
  \centering
   \includegraphics[width=0.9\linewidth]{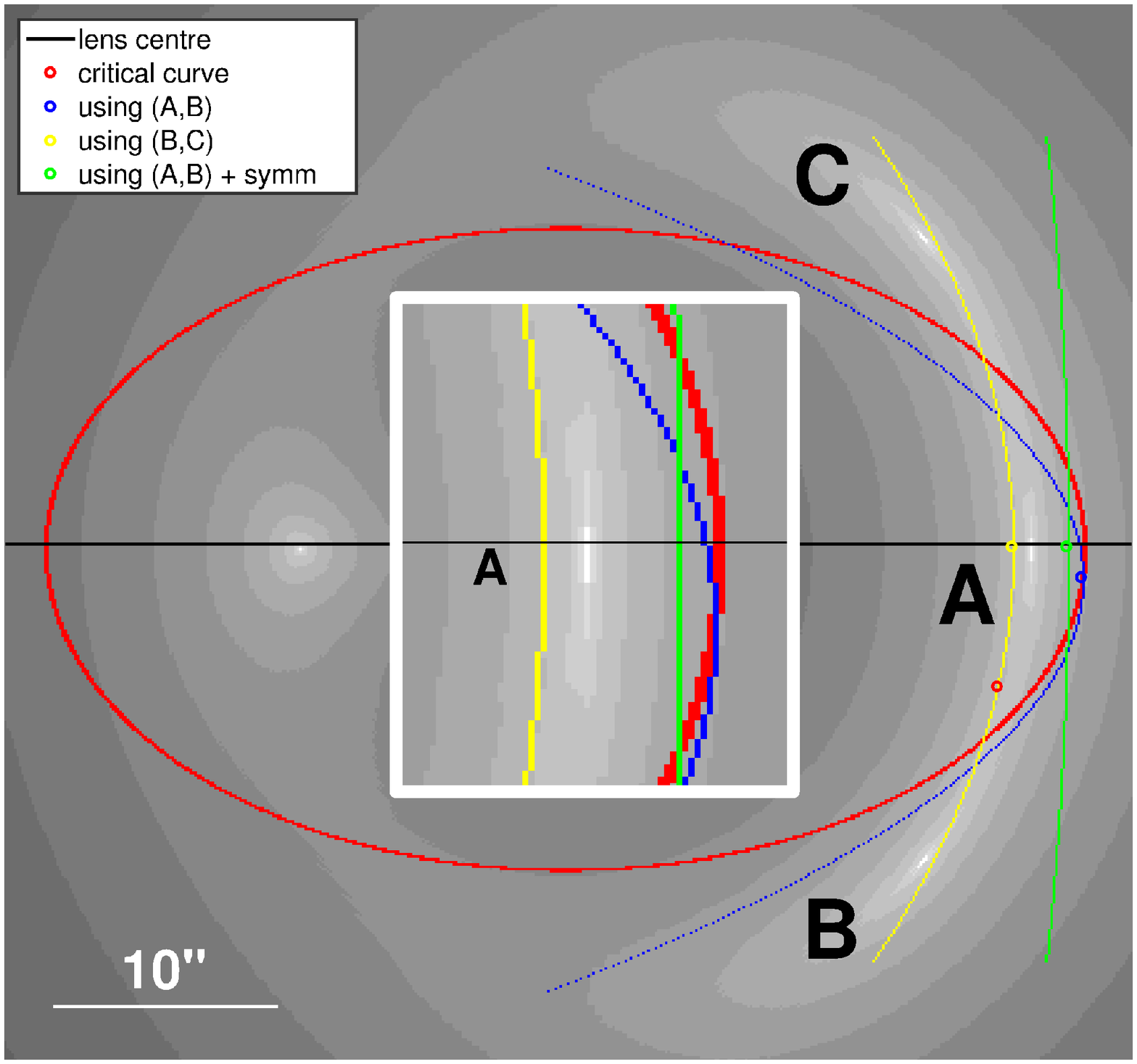}
  \caption{}
  \end{subfigure}
  \begin{subfigure}{0.495\textwidth}
  \centering
   \includegraphics[width=0.9\linewidth]{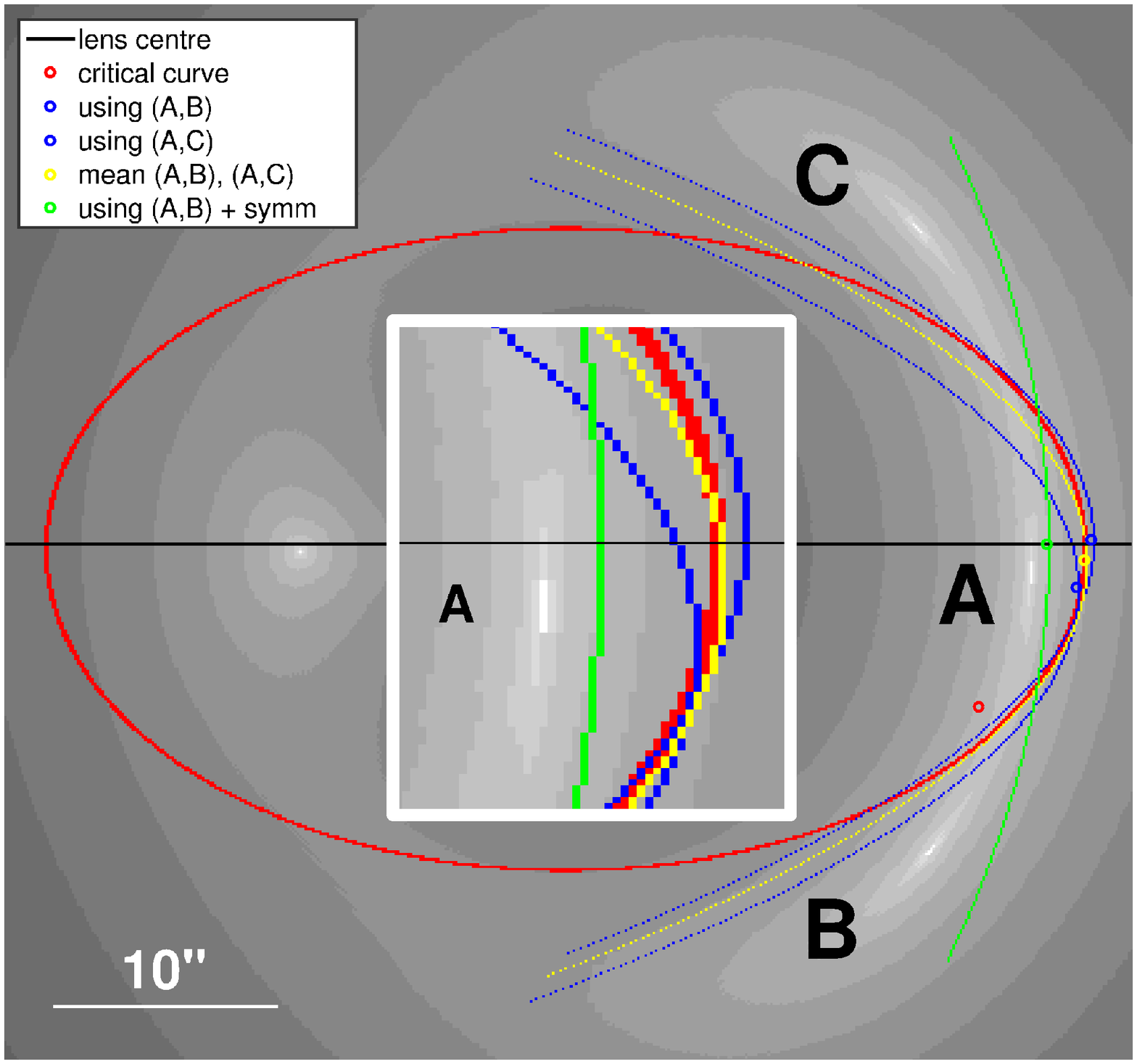}
  \caption{}
  \end{subfigure}
    \caption{Evaluation of the SIE-profile (cusp): Reconstruction of the critical curve close to a cusp for a symmetric image configuration (a) and for an asymmetric image configuration (b). For (a) and (b), the $y_1$-value of the sources is the same, $y_2$ varied. The symmetry axis through the lens centre (black line) is drawn to better visualise the location of the cusp point (crossing of black line and red curve) and compare it to the vertex positions (blue, yellow and green dots) for the different reconstructions, marking the reconstructed cusp point. The reconstruction of the fold point between $A$ and $B$ is also shown (red dot). On top, a detail of the cusp position reconstruction is shown (white square).}
 \label{fig:asymmetric_cusp}
\end{figure*}

At last, we investigate the difference between a symmetric and an asymmetric three-image configuration in the SIE lens. Fig.~\ref{fig:asymmetric_cusp} (a) and (b) show the image configurations for the symmetric and the asymmetric case, respectively. As can be read off (a), the approximation of the critical point under the symmetry assumption that the semi-major axes of the images are oriented parallel to the critical curve at the cusp is more accurate than the other approximations shown. However, the slope of the parabola is highly underestimated in this approach. A better result concerning the slope is obtained when reconstructing the critical curve using images $B$ and $C$ including their orientation angles, while the most accurate approximation of the slope is obtained using images $A$ and $B$ (or $A$ and $C$) including their orientation angles. The same results are obtained for the case, when the source lies off-axis and the image configuration becomes asymmetric. As can be observed in (b), the reconstruction of the critical point under the symmetry assumption is worse compared to (a) and compared to using $(A,B$) or $(A,C$).

The approximation of the cusp point using $(A,B)$ or $(A,C)$ is systematically biased towards the positive parity image used, $B$ or $C$. Due to the dependence of Eqs.~\eqref{eq:cusp_coords1} and \eqref{eq:cusp_coords2} on the magnification ratios $\mu_{AI} = \mu_I / \mu_A$, $I=A,B$, we calculate 
\begin{equation}
\boldsymbol{x}_0 = \dfrac{|\mu_{AC}| \left(\boldsymbol{x}_0\right)_{(A,B)} +|\mu_{AB}| \left(\boldsymbol{x}_0\right)_{(A,C)} }{|\mu_{AB}| + |\mu_{AC}|}
\end{equation}
denoted as the mean of the cusp position reconstructed from $(A,B)$ and $(A,C)$ and obtain that, in the symmetric case, the mean yields the most accurate approximation of the critical curve, while in the asymmetric case, the approximation using $(A,C)$ is most accurate.

Comparing Fig.~\ref{fig:asymmetric_cusp} (b) with the theoretical statements of Sect.~\ref{sec:cusp_fold}, we verify that the fold critical point obtained from the evaluation of $(A,B)$ as fold configuration lies within the parabolic approximation of the critical curve obtained from $(A,B)$ as part of a cusp configuration. In this case, $d(\boldsymbol{x}_A, \boldsymbol{x}_\mathrm{c}) \approx 3'' < d(\boldsymbol{x}_A, \boldsymbol{x}_\mathrm{f}) \approx 6.5''$, so that we expect the fold reconstruction to be less accurate than the cusp one, according to Sect.~\ref{sec:cusp_fold}. Using the orthogonality assumption, we observe that the reconstruction of the fold position and the slope of the critical curve achieves the same level of accuracy as the reconstruction of the critical curve for the nearby cusp. Hence, it is possible to build a fold configuration from the images with the shortest distance in a cusp configuration and thereby obtain additional information on the properties of the critical curve at the fold point. This also implies that a cusp configuration in which image $C$ is not observable cannot be distinguished from a fold configuration.

%%%%%%%%%%%%%%%%%
\section{Conclusion}
\label{sec:conclusion}

In this paper, we set the theoretical foundations of the model-independent characterisation of strong gravitational lenses by using the complete set of observables that can be employed in the leading order approach and generalised the equations developed in \cite{bib:Wagner}. We showed that the latter are retrieved from the generalised equations by employing symmetry assumptions, neglecting the information contained in the orientation angles of the images. We extended the analysis of biases started in \cite{bib:Wagner} for multiple images of point-like sources in elliptical lens models with multiple images of extended sources approximated by the quadrupole moment of their intensity distribution. Furthermore, we found that the critical point and the slope of the critical curve can even be determined in cases of overlapping images and that the quadrupole moment is a well-suited observable even in cases of non-elliptical isocontours.

Analysing the bias introduced by a Taylor-expanded lensing potential in the vicinity of a point on the critical curve, we conclude the following:
\begin{itemize}
\item The deviation in the reconstruction of fold and cusp critical points is below 5'' for multiple images closer than 30\% of the (effective) Einstein radius in a cluster scale NFW- and SIE-lens.
\item Breaking the mass-sheet-degeneracy at a reasonable accuracy in such lenses can only be achieved in a small range of distance between the images close to a cusp using the symmetry assumptions stated in \cite{bib:Wagner}.
\item Using the symmetry assumption to retrieve $\phi_{222}^{(0)}/\phi_{11}^{(0)}$ at a fold and including the orientation angles of the images to obtain $\phi_{122}^{(0)}/\phi_{11}^{(0)}$, the ratios of potential derivatives and the slope of the critical curve at a fold are reconstructed with accuracies above 85\% for relative image distances over 40\% of the (effective) Einstein radius in the NFW- and the SIE-lens.
\item Close to a cusp in the SIE-lens, including the orientation angles yields more accurate ratios of potential derivatives and the symmetry assumption of \cite{bib:Wagner} is only superior for the reconstruction of the cusp point from an image configuration whose source lies on the symmetry axis connecting the lens centre and the cusp in the caustic. The most accurate reconstruction of the properties of the critical curve at a cusp by a general (asymmetric) image configuration is obtained using the central image close to the cusp and the image farthest away from it. $\phi_{122}^{(0)}/\phi_{11}^{(0)}$ is reconstructed with an accuracy above 86\% for distances of the central image to the cusp up to 22\% of the effective Einstein radius. $\phi_{2222}^{(0)}/\phi_{11}^{(0)}$ has a much lower accuracy in the range of 75\%, together yielding an accuracy above 80\% for the slope, if the central image is closer to the cusp than 15\% of the effective Einstein radius. 
\item In a general (asymmetric) image configuration at a cusp, the two images closest to each other can be treated as a fold configuration and additional information on the slope and the position of the fold point can be obtained with the same level of accuracy as for the cusp.
\end{itemize}

Next, we plan to analyse multiple images in the HERA cluster, \cite{bib:Meneghetti}, in order to test the applicability of our approach in a realistic galaxy cluster scale simulation. Using this simulation, we will also investigate the uncertainty of the (ratios of) potential derivatives and the critical point positions due to observational imprecisions (noise and the convolution by the point-spread-function) and the validity of the assumption that intrinsic source properties are negligible.

%%%%%%%%%%%%%%%%%%%%%%%%%%%%%%%%%%%%%%%%%%%%%%%%%%%%%%%%%%%%%%%%%%%%%%%%%%%
\begin{acknowledgements}
I am grateful to Matthias Bartelmann and R\"{u}diger Vaas for their strong support. I wish to thank Mauricio Carrasco, Agnese Fabris, Matteo Maturi, Sven Meyer, Bj\"{o}rn-Malte Sch\"afer, Gregor Seidel, Nicolas Tessore, and the anonymous referee for helpful comments and discussions and gratefully acknowledge the support by the Deutsche Forschungsgemeinschaft (DFG) WA3547/1-1.\end{acknowledgements}

\bibliographystyle{aa}
\bibliography{aa}

\begin{thebibliography}{12}
\expandafter\ifx\csname natexlab\endcsname\relax\def\natexlab#1{#1}\fi

\bibitem[{{Bartelmann}(1996)}]{bib:Bartelmann2}
{Bartelmann}, M. 1996, Astronomy and Astrophysics, 313, 697

\bibitem[{{Bartelmann} \& {Schneider}(2001)}]{bib:Bartelmann}
{Bartelmann}, M. \& {Schneider}, P. 2001, \physrep, 340, 291

\bibitem[{{Bertin} \& {Arnouts}(1996)}]{bib:Bertin}
{Bertin}, E. \& {Arnouts}, S. 1996, \aaps, 117, 393

\bibitem[{Graham \& Driver(2005)}]{bib:Sersic}
Graham, A.~W. \& Driver, S.~P. 2005, Publications of the Astronomical Society
  of Australia, 22, 118

\bibitem[{{Kaiser} {et~al.}(1995){Kaiser}, {Squires}, \&
  {Broadhurst}}]{bib:KSB}
{Kaiser}, N., {Squires}, G., \& {Broadhurst}, T. 1995, \apj, 449, 460

\bibitem[{{Meneghetti} {et~al.}(2016){Meneghetti}, {Natarajan}, {Coe},
  {Contini}, {De Lucia}, {Giocoli}, {Acebron}, {Borgani}, {Bradac}, {Diego},
  {Hoag}, {Ishigaki}, {Johnson}, {Jullo}, {Kawamata}, {Lam}, {Limousin},
  {Liesenborgs}, {Oguri}, {Sebesta}, {Sharon}, {Williams}, \&
  {Zitrin}}]{bib:Meneghetti}
{Meneghetti}, M., {Natarajan}, P., {Coe}, D., {et~al.} 2016, ArXiv e-prints
  [\eprint[arXiv]{1606.04548}]

\bibitem[{{Navarro} {et~al.}(1996){Navarro}, {Frenk}, \& {White}}]{bib:NFW}
{Navarro}, J.~F., {Frenk}, C.~S., \& {White}, S.~D.~M. 1996, \apj, 462, 563

\bibitem[{Petters {et~al.}(2001)Petters, Levine, \& Wambsganss}]{bib:Petters}
Petters, A.~O., Levine, H., \& Wambsganss, J. 2001, Singularity Theory and
  Gravitational Lensing, Progress in Mathematical Physics, Volume 21
  (Birkh\"{a}user)

\bibitem[{{Refregier} \& {Bacon}(2003)}]{bib:Refregier}
{Refregier}, A. \& {Bacon}, D. 2003, \mnras, 338, 48

\bibitem[{Schneider {et~al.}(1992)Schneider, Ehlers, \& Falco}]{bib:SEF}
Schneider, P., Ehlers, J., \& Falco, E.~E. 1992, {Gravitational Lenses},
  {Astronomy and Astrophysics Library} (New York: Springer)

\bibitem[{{Seitz} \& {Schneider}(1995)}]{bib:Seitz}
{Seitz}, C. \& {Schneider}, P. 1995, \aap, 297, 287

\bibitem[{{Wagner, J.} \& {Bartelmann, M.}(2016)}]{bib:Wagner}
{Wagner, J.} \& {Bartelmann, M.} 2016, Astronomy \& Astrophysics, 590, A34

\end{thebibliography}

%%%%%%%%%%%%%%%%%%%%%%%%%%%%%%%%%%%%%%%%%%%%%%%%%%%%%%%%%%%%%%%%%%%%%%%%%%%%
\appendix
\section{Influence of source properties}
\label{app:source_properties}

To show the influence of the source properties on the image observables, we use Eq.~(4.12) from \cite{bib:Bartelmann} as a starting point
\begin{equation}
\epsilon_\mathrm{s} = \dfrac{\epsilon - g}{1 - g^* \epsilon} \;, \quad |g| \le 1 \;,
\end{equation}
in which $\epsilon$ and $\epsilon_\mathrm{s}$ are the complex variables representing the two-dimensional ellipticities of the image(s) and the source, respectively, and $g$ is the complex variable representing the two-dimensional reduced shear
\begin{equation}
g = \dfrac{\gamma}{1-\kappa} = \dfrac{\gamma_1 + \mathrm{i} \gamma_2}{1-\kappa} \;.
\end{equation}
In the following, we will derive the influence of the source properties for $|g| \le 1$. The case $|g| > 1$ is analogous.
Expressing the shear components $\gamma_1$ and $\gamma_2$ and the convergence $\kappa$ in potential derivatives, we obtain
\begin{equation}
\gamma_1 = \dfrac12 \left( \phi_{22} - \phi_{11} \right)\;, \quad \gamma_2 = \phi_{12}\;, \quad \kappa = 2\left( 1- \phi_{11}\right)
\end{equation}
such that $|g| = 1$ at the critical curve. Transforming the complex variables $\epsilon, \epsilon_s, g$ into their polar representation, we can express the absolute value and the argument of $\epsilon$ as
\begin{eqnarray}
|\epsilon| &= |g| \; |z| \;, \; &\text{with} \quad |z| = |z(\epsilon_\mathrm{s}, g)| \propto |g|^{-1} \propto x_2^{-1} \;, \\
\varphi_\epsilon &= \varphi_g + \tfrac12 \varphi_z \;, \; &\text{with} \quad \varphi_z = \varphi_z(\epsilon_\mathrm{s}, g) \propto |g|^2-1 \propto x_2^2 \;.
\end{eqnarray}
Hence, as $|g| \rightarrow 1$ and $x_2 \rightarrow 0$, 
\begin{equation}
|\epsilon| \rightarrow 1 \;, \quad \Leftrightarrow \quad r = \dfrac{1-|\epsilon|}{1+|\epsilon|} \rightarrow 0 \;.
\end{equation}
In the same limit, the argument $\varphi_\epsilon$, which also represents the orientation angle of the semi-major axis of an image becomes aligned with the orientation of the shear, $\varphi_g$, as $\varphi_z \rightarrow 0$.

%%%%%%%%%%%%%%%%%%
\section{Derivation for the fold case}
\label{app:fold_derivation}
For the following derivations, we rely on Chapter~6 of \cite{bib:SEF}. Setting the derivative of the Taylor-expanded lensing potential to zero, we arrive at the lensing equations
\begin{align}
  y_1 &= \phi_{11}^{(0)}x_{1} + \dfrac12\phi_{122}^{(0)}x_{2}^2 + \phi_{112}^{(0)}x_{1}x_{2}\;,
  \label{eq:fold_le1} \\
  y_2 &= \dfrac12\phi_{112}^{(0)}x_{1}^2 + \phi_{122}^{(0)}x_{1}x_{2} + \dfrac12\phi_{222}^{(0)}x_{2}^2\;.
  \label{eq:fold_le2}
\end{align}
To leading order, \cite{bib:SEF} obtain that two images $A$ and $B$ close to the critical curve have the same $x_1$-coordinate and their distance to the critical curve, measured along $x_2$, is half of their mutual distance. Employing these results and the fact that $A$ and $B$ have the same source coordinates $(y_1, y_2)$, we eliminate the latter and, without loss of generality, $\boldsymbol{x}_B$ to obtain
\begin{align}
  0 &= \phi_{112}^{(0)}x_{A1} + \phi_{122}^{(0)}x_{A2} - \dfrac12 \phi_{122}^{(0)}\delta_{AB2}\;,
  \label{eq:fold_nos_le1} \\
  0 &= \phi_{122}^{(0)}x_{A1} + \phi_{222}^{(0)}x_{A2} - \dfrac12\phi_{222}^{(0)}\delta_{AB2}\;.
  \label{eq:fold_nos_le2}
\end{align}
To eliminate the unknown $x_{A1}$ and $x_{A2}$, we insert the Taylor expansions
\begin{align}
\phi_{12}^{(A)} &\approx  \phi_{112}^{(0)}x_{A1} + \phi_{122}^{(0)}x_{A2} \;, \label{eq:fold_dm_le1} \\
\phi_{22}^{(A)} &\approx  \phi_{122}^{(0)}x_{A1} + \phi_{222}^{(0)}x_{A2} \label{eq:fold_dm_le2} 
\end{align}
for the first two terms on the right hand sides of Eqs.~\eqref{eq:fold_nos_le1} and \eqref{eq:fold_nos_le2}
\begin{align}
  0 &= \phi_{12}^{(A)} - \dfrac12 \phi_{122}^{(0)}\delta_{AB2}\;,
  \label{eq:fold_dist_le1} \\
  0 &= \phi_{22}^{(A)} - \dfrac12\phi_{222}^{(0)}\delta_{AB2}\;.
  \label{eq:fold_dist_le2}
\end{align}
Then, we connect the distortion matrix of $A$ with the quadrupole moment obtained from the image observables $r_A$ and $\varphi_A$. As $r_A$ and $\varphi_A$ do not depend on a (fixed) overall length scale, but the potential derivatives are still subject to that degeneracy, we scale them by $\phi_{11}^{(A)} \approx \phi_{11}^{(0)}$, which yields
\begin{align}
\dfrac{\phi_{12}^{(A)}}{\phi_{11}^{(0)}} &= \dfrac{\lambda_A (r_A-1)}{r_A \lambda_A^2 + 1} \;, \label{eq:fold_dist1} \\
\dfrac{\phi_{22}^{(A)}}{\phi_{11}^{(0)}} &= \dfrac{r_A + \lambda_A^2}{r_A \lambda_A^2 + 1} \;. \label{eq:fold_dist2}
\end{align}
In this step, we assume that both images are close enough to the critical curve that intrinsic ellipticities are negligible compared to the distortions caused by lensing.
Inserting Eqs.~\eqref{eq:fold_dist1} and \eqref{eq:fold_dist2} into  Eqs.~\eqref{eq:fold_dist_le1} and \eqref{eq:fold_dist_le2}, we arrive at Eqs.~\eqref{eq:fold2} and \eqref{eq:fold3}.

The coordinates of the images with respect to the critical curve are obtained by the entries of the distortion matrix and Eqs.~\eqref{eq:fold_dm_le1} and \eqref{eq:fold_dm_le2}.
\begin{align}
\dfrac{\phi_{12}^{(A)}}{\phi_{11}^{(0)}} &= \dfrac{\lambda_A (r_A-1)}{r_A \lambda_A^2 + 1} =  \dfrac{\phi_{122}^{(0)}}{\phi_{11}^{(0)}} x_{A2} \;, \\
\dfrac{\phi_{22}^{(A)}}{\phi_{11}^{(0)}} &= \dfrac{r_A + \lambda_A^2}{r_A \lambda_A^2 + 1} = \dfrac{\phi_{122}^{(0)}}{\phi_{11}^{(0)}} x_{A1} + \dfrac{\phi_{222}^{(0)}}{\phi_{11}^{(0)}} x_{A2} \;,
\end{align} 
inserting Eqs.~\eqref{eq:fold2} and \eqref{eq:fold3} for the ratios of potential derivatives on the right hand sides.

%%%%%%%%%%%%%%%%%%
\section{Derivation for the cusp case}
\label{app:cusp_derivation}

Analogously to Appendix~\ref{app:fold_derivation}, the ratios of potential derivatives of the cusp case can be retrieved. The lensing equations of the Taylor-expanded potential read
\begin{align}
  y_1 &= \phi_{11}^{(0)}x_1 + \dfrac{1}{2} \phi_{122}^{(0)}x_2^2\;, \label{eq:cusp_le1} \\
  y_2 &= \phi_{122}^{(0)}x_1x_2 + \dfrac{1}{6}\phi_{2222}^{(0)}x_2^3\;.
\label{eq:cusp_le2}
\end{align}
Eliminating the source coordinates and using the Taylor expansions
\begin{align}
\phi_{12}^{(A)}  &\approx  \phi_{122}^{(0)}x_{A2} \;, \label{eq:cusp_Taylor1} \\
\phi_{22}^{(A)} &\approx \phi_{122}^{(0)}x_{A1} + \dfrac12 \phi_{2222}^{(0)} x_{A2}^2 \label{eq:cusp_Taylor2}\;, 
\end{align}
we arrive at
\begin{align}
  0 &= \phi_{11}^{(0)} \delta_{AB1} +  \left( \phi_{12}^{(A)}- \dfrac12  \phi_{122}^{(0)}\delta_{AB2}\right) \delta_{AB2}\;,
  \label{eq:cusp_dist_le1} \\
  0 &= \left( \phi_{12}^{(A)}- \phi_{122}^{(0)}\delta_{AB2}\right) \delta_{AB1} + \left( \phi_{22}^{(A)} + \dfrac16 \phi_{2222}^{(0)} \delta_{AB2}^2 \right) \delta_{AB2}\;,
  \label{eq:cusp_dist_le2}
\end{align}
if we also assume that the flexion term $\phi_{222}^{(A)} \delta_{AB2}$ is small compared to the other terms in the last bracket on the right hand side of Eq.~\eqref{eq:cusp_dist_le2} and can thus be neglected. Replacing $B$ by $C$, we obtain the lensing equations for the image pair $A$ and $C$. Replacing the second order potential derivatives at the image positions by Eqs.~\eqref{eq:fold_dist1} and \eqref{eq:fold_dist2} (or its negative in case of a positive cusp configuration), Eqs.\ref{eq:cusp_dist_le1} and \ref{eq:cusp_dist_le2} yield Eqs.~\eqref{eq:cusp2} and \eqref{eq:cusp3}.

From Eq.~\eqref{eq:cusp_dist_le1}, we determine the angle that $\boldsymbol{\delta}_{AB}$ has with the $x_1$-axis of the cusp coordinate system, $\alpha$, such that we can express $\delta_{Ai1}$ and $\delta_{Ai2}$, $i=B,C$ as
 \begin{align}
\delta_{AB1} =& -|\boldsymbol{\delta}_{AB}| \, \cos\left(\alpha\right) < 0 \;, &\delta_{AB2} =& -|\boldsymbol{\delta}_{AB}| \, \sin \left(\alpha\right) < 0\;, \label{eq:cusp_trigonometry1}\\
\delta_{AC1} =& -|\boldsymbol{\delta}_{AC}| \, \cos\left(\alpha_A - \alpha\right) < 0\;, &\delta_{AC2} =& \phantom{-} |\boldsymbol{\delta}_{AC}| \, \sin \left(\alpha_A - \alpha\right)> 0 \;,\label{eq:cusp_trigonometry2}
\end{align}
where we also used the opening angle between $\boldsymbol{\delta}_{AB}$ and $\boldsymbol{\delta}_{AC}$, denoted as $\alpha_A$. 

$\alpha$ is determined by first inserting $\phi_{12}^{(A)} - \phi_{12}^{(B)} = \phi_{122}^{(0)} \delta_{AB2}$ from Eq.~\eqref{eq:cusp_Taylor1} into Eq.~\eqref{eq:cusp_dist_le1}, replacing $\delta_{AB1}$ and $\delta_{AB2}$ by Eq.~\eqref{eq:cusp_trigonometry1} and then numerically solving the resulting equation 
\begin{equation}
\dfrac{-2}{\tan(\alpha)} = \dfrac{(r_A-1) \tan \left( \alpha + \tilde{\varphi}_A \right)}{r_A + \tan^2 \left( \alpha + \tilde{\varphi}_A \right)} + \dfrac{(r_B-1) \tan \left( \alpha + \tilde{\varphi}_B \right)}{r_B + \tan^2 \left( \alpha + \tilde{\varphi}_B \right)} \;,
\end{equation}
where the orientation angles between the semi-major axis of images $A$ and $B$ are measured with respect to $\boldsymbol{\delta}_{AB}$ and called $\tilde{\varphi}_A$ and $\tilde{\varphi}_B$, respectively (s.\ Fig.~\ref{fig:coordinate_systems} (b)). 

For the coordinates of $\boldsymbol{x}_A$, we use the flux ratio that is equal to the inverse of the respective ratio of distortion matrix determinants. Expressing the determinant of $B$ by that of $A$ and scaling all potential derivatives by $\phi_{11}^{(0)}$ yields
\begin{align}
\dfrac{\mu_A}{\mu_B} = 1 - \dfrac{\tfrac{\phi_{122}^{(0)}}{\phi_{11}^{(0)}} \delta_{AB1} + \left(\tfrac12 \tfrac{\phi_{2222}^{(0)}}{\phi_{11}^{(0)}} - \left( \tfrac{\phi_{122}^{(0)}}{\phi_{11}^{(0)}} \right)^2 \right)(2 x_{A2} - \delta_{AB2})\delta_{AB2}}{\det(\tilde{M}_A)}
\label{eq:flux_ratio}
\end{align}
with $\det(\tilde{M}_A)$ given by Eq.~\eqref{eq:det_MA}. Solving Eq.~\eqref{eq:flux_ratio} for $x_{A2}$ yields Eq.~\eqref{eq:cusp_coords2}. $x_{A1}$ in Eq.~\eqref{eq:cusp_coords1} is then obtained from combining Eq.~\eqref{eq:fold_dist2} (or its negative in the case of a positve cusp configuration) with Eq.~\eqref{eq:cusp_Taylor2} scaled by $\phi_{11}^{(0)}$
\begin{equation}
\dfrac{\phi_{22}^{(A)}}{\phi_{11}^{(0)}} = \dfrac{\phi_{122}^{(0)}}{\phi_{11}^{(0)}} x_{A1} + \dfrac12 \dfrac{\phi_{2222}^{(0)}}{\phi_{11}^{(0)}} x_{A2}^2 \;.
\end{equation}

The time-delay is calculated from the difference of the lensing potentials $\Delta \phi_{AB} = \phi(\boldsymbol{y},\boldsymbol{x}_A)-\phi(\boldsymbol{y},\boldsymbol{x}_B)$. To leading order in the Taylor expansion this potential difference reads
\begin{align}
\Delta \phi_{AB} =& -y_1 \delta_{AB1} - y_2 \delta_{AB2} + \dfrac12 \phi_{11}^{(0)} \left( x_{A1}^2 - x_{B1}^2 \right) \nonumber \\
&+ \dfrac12 \phi_{122}^{(0)} \left( x_{A1} x_{A2}^2 - x_{B1} x_{B2}^2 \right) + \dfrac{1}{24} \phi_{2222}^{(0)} \left( x_{A2}^4 - x_{B2}^4 \right) \;.
\end{align}
Inserting Eqs.~\eqref{eq:cusp_le1} and \eqref{eq:cusp_le2}, subsequently Eqs.~\eqref{eq:cusp_Taylor1} and \eqref{eq:cusp_Taylor2} and Eqs.~\eqref{eq:fold_dist1} and \eqref{eq:fold_dist2}, we arrive at Eq.~\eqref{eq:cusp3}, noting that $t = \Delta \phi \Gamma_\mathrm{d}/c$. 

%%%%%%%%%%%%%%%%%%%
\section{Derivation of Eq.~\eqref{eq:delta_x}}
\label{app:fold_cusp_comparison}

As shown in \cite{bib:SEF}, placing a source at $\boldsymbol{y} = (y_\mathrm{s},0)$, its image positions of $A$ and $B$ in the coordinate system of the cusp configuration are given by 
\begin{align}
\boldsymbol{x}_A &= \left( \dfrac{y_\mathrm{s}}{\phi_{11}^{(0)}}, 0 \right) \;, \\
\boldsymbol{x}_B &= \left(  \dfrac{\phi_{2222}^{(0)} y_\mathrm{s}}{\phi_{11}^{(0)}\phi_{2222}^{(0)}- 3 \left( \phi_{122}^{(0)}\right)^2}, \sqrt{\dfrac{-6\phi_{122}^{(0)} y_\mathrm{s}}{\phi_{11}^{(0)}\phi_{2222}^{(0)}-3 \left(\phi_{122}^{(0)} \right)^2}}\right) \;,
\end{align}
so that their enclosed fold critical point is located at
\begin{equation}
\boldsymbol{x}_\mathrm{f} = \left( \dfrac{y_\mathrm{s} \left( 2\phi_{11}^{(0)}\phi_{2222}^{(0)}-3 \left(\phi_{122}^{(0)} \right)^2\right)}{2\phi_{11}^{(0)} \left( \phi_{11}^{(0)}\phi_{2222}^{(0)}-3 \left(\phi_{122}^{(0)} \right)^2\right) },\sqrt{\dfrac{-3\phi_{122}^{(0)} y_\mathrm{s}} {2\left( \phi_{11}^{(0)}\phi_{2222}^{(0)}-3 \left(\phi_{122}^{(0)} \right)^2\right) }} \right) \;.
\label{eq:x2_fold}
\end{equation}
The approximation of the critical curve as a parabola along the $x_1$-axis reads
\begin{equation}
x_1 = m_\mathrm{c} x_2^2
\end{equation}
using $m_\mathrm{c}$ of Eq.~\eqref{eq:parabola_slope}, such that the $x_2$-coordinate at $x_{\mathrm{f}1}$ reads
\begin{equation}
x_{\mathrm{c}2} = \sqrt{\dfrac{-\phi_{122}^{(0)} y_\mathrm{s}\left( 2 \phi_{11}^{(0)}\phi_{2222}^{(0)}-3 \left(\phi_{122}^{(0)} \right)^2\right)} {\left( \phi_{11}^{(0)}\phi_{2222}^{(0)}-3 \left(\phi_{122}^{(0)} \right)^2\right)  \left( \phi_{11}^{(0)}\phi_{2222}^{(0)}-2 \left(\phi_{122}^{(0)} \right)^2\right)} } \;.
\label{eq:x2_cusp}
\end{equation}
Employing $x_{\mathrm{f}2}$ of Eq.~\eqref{eq:x2_fold} and Eq.~\eqref{eq:x2_cusp} yields the expression for $\Delta x_2 = x_{\mathrm{c}2} -x_{\mathrm{f}2} $ of Eq.~\eqref{eq:delta_x}.

%%%%%%%%%%%%%%%%%%%%%%%%%%%%%%%%%%%%%%%%%%%%%%%%%%%%%%%%%%%%%%%%%%%%%%%%%%%%%
\end{document}